\documentclass[11pt]{article}
\newcommand{\Comment}[1]{{}}
\usepackage[textwidth = 430 pt, textheight = 630 pt]{geometry}
\usepackage{amssymb,euscript,amsmath,amsfonts,filecontents}
\usepackage[dvipsnames]{xcolor}
\usepackage[utf8]{inputenc}

\usepackage{color}
\definecolor{MyDarkBlue}{rgb}{0.15,0.15,0.45}
\usepackage[linktocpage=true]{hyperref}
\hypersetup{
colorlinks=true,
citecolor=MyDarkBlue,
linkcolor=MyDarkBlue,
urlcolor=MyDarkBlue,
pdfauthor={E. Andriolo, N. Lambert and A. Papageorgakis},
pdftitle={Geometrical Aspects of an Abelian (2,0) Action},
pdfsubject={hep-th}
}

\usepackage[numbers,sort&compress]{natbib}
\usepackage{hypernat}

\newcommand{\ret}{\nonumber \\}
\newcommand{\nn}{\nonumber}

\newcommand{\be}{\begin{equation}}
\newcommand{\ee}{\end{equation}}
\newcommand{\bea}{\begin{eqnarray}}
\newcommand{\eea}{\end{eqnarray}}

\numberwithin{equation}{section}

\begin{document}

\renewcommand{\thefootnote}{\fnsymbol{footnote}}
 
 \rightline{QMUL-PH-20-06}

   \vspace{1.8truecm}

 \centerline{\LARGE \bf {\sc Geometrical Aspects of An Abelian (2,0) Action}}
 
 \vspace{1cm}
 
   \centerline{
   {\large {\bf {\sc E.~Andriolo${}^{\,a}$}\footnote{E-mail address: \href{e.andriolo@qmul.ac.uk}{\tt e.andriolo@qmul.ac.uk}}\,,  {\sc N.~Lambert${}^{\,b,c}$}}\footnote{E-mail address: \href{neil.lambert@kcl.ac.uk}{\tt neil.lambert@kcl.ac.uk}}\,   {\sc    and C.~Papageorgakis${}^{\,a}$}}\footnote{E-mail address: \href{mailto:c.papageorgakis@qmul.ac.uk}{\tt c.papageorgakis@qmul.ac.uk}}  }

\vspace{1cm}
\centerline{${}^a${\it CRST and School of Physics and Astronomy }}
\centerline{{\it Queen Mary University of London,  E1 4NS, UK}} 
\vskip 1cm
\centerline{${}^b${\it Department of Mathematics}}
\centerline{{\it King's College London, WC2R 2LS, UK}} 
\vskip 1cm
\centerline{${}^c${\it Department of Theoretical Physics, CERN}}
\centerline{{\it CH-1211 Geneva, Switzerland}}

\vspace{1.0truecm}

 
\thispagestyle{empty}

\centerline{\sc Abstract}
\vspace{0.4truecm}
\begin{center}
\begin{minipage}[c]{360pt}{We explore various geometrical aspects of an action for six-dimensional chiral 2-forms based on the formalism  of 1903.12196. We elucidate the coupling to general backgrounds and construct the full supersymmetric completion to an abelian $(2,0)$ superconformal lagrangian including matter. We investigate the non-standard diffeomorphism properties of the fields and their relation to the hamiltonian formulation. We also test the action by considering compactifications  on a circle, K3 and a Riemann surface. The results are consistent with expectations for an action describing the low-energy physics of an M5-brane in M-theory.
    \noindent}
 
\end{minipage}
\end{center}
\renewcommand{\thefootnote}{\arabic{footnote}}
\setcounter{footnote}{0}
\newpage 
\tableofcontents

\section{Introduction}\label{sect: Intro}
Over the years there have been many approaches to construct a lagrangian that captures the low-energy dynamics of M5-branes in M-theory. In that course, a number of arguments have emerged strongly suggesting that the interacting theory is inherently non-lagrangian; for an executive summary see \cite{Lambert:2019diy}. Indeed, even the construction of a lagrangian describing the low-energy dynamics of a single M5-brane is non-trivial, although Lorentz invariant supersymmetric equations of motion have been constructed to all orders in \cite{Howe:1997fb}. This is due to the physical spectrum being encoded in a free (2,0) tensor multiplet, containing a chiral 2-form.

There are well-known difficulties in writing down lagrangians for theories involving chiral $2k$-forms in $4k+2$ dimensions \cite{Henneaux:1988gg}. An initial work-around involved imposing the self-duality condition by hand at the level of the equations of motion, after deriving the latter from an action. Subsequently, various alternative formulations emerged where the self-duality condition arises on-shell, but at the cost of either breaking manifest Lorentz invariance \cite{Henneaux:1988gg,Perry:1996mk,Aganagic:1997zq}, introducing an infinite number of auxiliary fields \cite{McClain:1990sx,Wotzasek:1990zr,Martin:1994np, Devecchi:1996cp, Faddeev:1986pc, Bengtsson:1996fm,Berkovits:1996nq,Berkovits:1996rt}, or requiring an extra dimension and considering a $4k+3$-dimensional Chern--Simons theory \cite{Belov:2006jd,Witten:1996hc,Witten:1999vg}. Last but not least, one can write a manifestly Lorentz-invariant action for chiral forms where the auxiliary fields are finite in number but enter in a non-polynomial way; this is the so-called Pasti--Sorokin--Tonin (PST) formalism  \cite{Pasti:1995us,Pasti:1996vs,Pasti:1997gx,Bandos:1997ui,DallAgata:1997gnw,DallAgata:1998ahf,DallAgata:1998rvz,Bandos:2014bva}. For other interesting recent work regarding chiral forms see \cite{Saemann:2011nb,Mason:2011nw,Huang:2018hho,Mkrtchyan:2019opf,Buratti:2019guq,Jurco:2019bvp,Saemann:2019dsl,Townsend:2019koy}.

Recently, Sen put forward a new string-field-theory inspired proposal for a lagrangian description of  chiral $2k$-forms in $4k+2$ dimensions \cite{Sen:2015nph} (see also \cite{Sen:2019qit}), where the self-duality condition holds off-shell. This deploys auxiliary degrees of freedom in a polynomial way, while preserving manifest Lorentz invariance. The invariance of the action under general diffeomorphisms is not manifest, because the coupling to gravity is realised in a non-standard fashion. Moreover, the action does not couple the fields to the metric in the usual covariant way and, therefore, there is room to evade the no-go theorems regarding the compactifications of chiral $2k$-forms actions \cite{Witten:2009at}. These attractive properties make this proposal worthy of further study.

In this paper, we will focus on the action for chiral 2-form given by \cite{Sen:2019qit}
\begin{align}
S_H= \int \left(\frac12 dB\wedge\star_\eta dB -2H\wedge dB +   H\wedge {\tilde {\cal M}}(H) \right)\ .
\end{align}
Here $B$ is a generic ``2-form'', while $H$ is a chiral ``3-form'' subject to the self-duality condition $H = \star_\eta H$. This expression has some unconventional features. For instance, the coupling to the background is achieved via the interaction term involving $\tilde {\mathcal M}$, which is a function of the metric $g$ only. We stress that, although the background is generically curved ($g\not=\eta$), the Hodge star entering the kinetic term of $B$ is defined with respect to the flat Minkowski metric. As a result, $B$ and $H$ are not standard differential forms, a fact that is also reflected in their non-standard transformation properties under diffeomorphisms.
It turns out that $S_H$ encodes on-shell the degrees of freedom carried by---not one but---two free 2-forms with self-dual field strength: in the hamiltonian formulation, it can be shown that the theory contains an unphysical sector (with a wrong-sign kinetic term) that explicitly decouples from the physical one \citep{Sen:2019qit}. Thus one expects the physical sector to correctly describe the physics of free chiral 2-forms on generic manifolds. 

The supersymmetric completion of this model to a $(2,0)$ theory for Minkowski space was constructed in \cite{Lambert:2019diy}. In this paper we further investigate and extend several aspects of this $(2,0)$ lagrangian. In Sec.~\ref{sect: SUSY} we first elucidate the nature of the coupling of the dynamical degrees of freedom to arbitrary backgrounds, providing an alternative to the perturbative construction of $\tilde {\mathcal M}$ given in \cite{Sen:2019qit}; we also discuss the introduction of sources. We then revisit the (non-manifest) diffeomorphism invariance of the theory and show that the action reproduces standard results following from diffeomorphism-invariant theories, by {\it e.g.} evaluating the energy-momentum tensor. This information allows us to identify two particular  combinations of the lagrangian fields $B$ and $H$
\begin{align}
H_{(s)}:=&H+\left(\frac{1+\star_\eta}{2}\right) dB\cr
H_{(g)}:=&H-\tilde {\mathcal M}(H)\;,
\end{align}
which respectively correspond (on shell) to a singlet ``3-form''  and a standard chiral 3-form under diffeomorphisms. We then re-examine the hamiltonian formulation of the theory and make apparent the fact that $H_{(s)}$ and $H_{(g)}$ are, respectively, the unphysical and physical chiral degrees of freedom of the theory. We also determine the hamiltonian in terms of $H_{(s)}$ and familiar geometric quantities such as the energy-momentum tensor of $H_{(g)}$. At the end of Sec.~\ref{sect: SUSY}, we provide an extension to the supersymmetric completion of the action for arbitrary backgrounds, that is for arbitrary $\tilde{\mathcal M}$.\footnote{One of the key inputs of \cite{Lambert:2019diy} was that the field $H_{(s)}$ is a singlet with respect to supersymmetry transformations.}

Then, in Sec.~\ref{compactifications}, we proceed to consider some applications and consistency checks of the action  by dimensionally reducing it on   a circle, K3 and a non-compact Riemann surface. The reductions are non-trivial and we use either the lagrangian or hamiltonian formulation on a case-by-case basis. The first example  leads to the expected spectra of a five-dimensional Maxwell theory, whose lagrangian scales inversely with the radius $R$, whereas the second example leads to the heterotic string transverse to  $\mathbb R^5\times \mathbb T^3$, plus some unphysical, decoupled degrees of freedom. The case of the Riemann surface is more interesting as the reduction depends on the scalars and hence is itself dynamical. We follow the approach of \cite{Howe:1997hxz, Lambert:1997dm} with the aim to reproduce the four-dimensional $\mathcal N=2$ Seiberg--Witten effective action \cite{Seiberg:1994rs}. We arrive at an action  for two---instead of one---sets of real, abelian gauge fields subject to a constraint that relates them via electric-magnetic duality. Furthermore, in this case the unphysical sector does not entirely decouple but rather  acts as a  background. We conclude with a summary and some open questions in Sec.~\ref{sec:conclusions}. 
 
\section{Abelian (2,0) Action on a Generic Manifold}\label{sect: SUSY}
 
We begin our discussion with a recap of the relevant background. In flat six-dimensional Minkowski spacetime one can write down the following action for the fields of the free (2,0) tensor multiplet \cite{Lambert:2019diy}
\begin{equation}\label{Sis}
S = \int  \ \left(
\frac12 dB\wedge\star dB -2H\wedge dB
- \frac12 \partial_\mu X^I\partial^\mu X^I  + \frac {i}{2}\bar\Psi\Gamma^\mu\partial_\mu \Psi\right)\ ,
\end{equation}
where  $H=\star_\eta H$.
This is invariant under the superconformal transformations 
\begin{align}\label{susyflat}
\delta X^I &=i\bar\epsilon \Gamma^I\Psi\ret
\delta B_{\mu\nu} & = -i\bar\epsilon\Gamma_{\mu\nu}\Psi\ret
\delta H_{\mu\nu\lambda} &  =   \frac{3i }{{2}} \bar\epsilon\Gamma_{[\mu\nu}\partial_{\lambda]}\Psi + \frac{3i }{{2}\cdot 3!}\varepsilon_{\mu\nu\lambda\rho\sigma\tau}\bar\epsilon\Gamma^{\rho\sigma}\partial^{\tau}\Psi - \frac{i}{2}\partial^\rho\bar\epsilon\Gamma_\rho\Gamma_{\mu\nu\lambda}\Psi\ret
\delta \Psi &= \Gamma^\mu\Gamma^I\partial_\mu X^I\epsilon+\frac{1}{  3!}\Gamma^{\mu\nu\lambda}H_{\mu\nu\lambda}\epsilon -\frac{2}{3}\Gamma^IX^I\Gamma^\rho\partial_\rho\epsilon\ ,
\end{align}
with
\begin{equation}
\partial_\mu \epsilon = \frac{1}{6}\Gamma_\mu\Gamma^\rho\partial_\rho\epsilon\ .
\end{equation}   
A key point of this system is that 
\begin{align}
	H_{(s)} = \frac12 (dB+\star dB)+H\ ,
\end{align}
 is a supersymmetry singlet and on-shell decouples from the rest of the fields. Of course, the latter statement is rather trivial as all fields are free and decoupled. But one can come up with interacting lagrangians for which $H_{(s)}$ is still decoupled.
 
It is desirable to extend this action to a general curved spacetime with metric $g$. In principle, one could easily try to couple it  in the usual way:
 \begin{equation}
 S = \int  \ \left(
\frac12 dB\wedge\star_g dB -2H\wedge dB
- \frac12 d X^I\wedge \star_g d X^I  + \frac {i}{2}\bar\Psi\Gamma_\mu dx^\mu \wedge \star_g\nabla \Psi - \frac15 RX^IX^I\right)\ ,
\end{equation}
where $R$ is the Ricci scalar,  $H = \star_g H$ with $\star_g$ the Hodge dual evaluated with respect to the metric $g$, and $\nabla$ is the corresponding covariant derivative on spinors.
Indeed this will still be supersymmetric if all expressions in \eqref{susyflat} are replaced with covariant ones and by assuming that there exists a spinor satisfying $\nabla_\mu\epsilon=\frac16\Gamma_\mu \Gamma^\rho\nabla_\rho\epsilon$.
However, this would imply that the spurious degrees of freedom associated with
$H_{(s)}$ also couple to the metric.

Rather, to make $B$ truly decoupled Sen \cite{Sen:2015nph,Sen:2019qit} considers the following
\begin{align}
S = S_H + S_{mat}	\ ,
\end{align}
where $S_{mat}$ is the usual action for the matter fields and $S_H$ is given by 
\begin{align}\label{SenAction}
S_H= \int \left(\frac12 dB\wedge\star_\eta dB -2H\wedge dB +   H\wedge {\tilde {\cal M}}(H) \right)\ ,
\end{align}
while still imposing the self-duality condition $H=\star_\eta H$.  
Here we have introduced a subscript on $\star_\eta $  to emphasise that, although the spacetime metric is nontrivial, the Hodge dual is evaluated with the flat Minkowski metric. This is not the expected behaviour for 3-forms on a nontrivial metric; we will in fact see in due course that this is reflected in their unusual transformation properties under diffeomorphisms. In the last term above, $\tilde {\cal M}$ is a linear map:
\begin{align}
  \tilde {\cal M}(H)_{\mu\nu\lambda} = \frac1{3!}\tilde {\cal M}_{\mu\nu\lambda}^{\alpha\beta\gamma}H_{\alpha\beta\gamma}\ .
\end{align}
Since only the anti-self-dual part of $\tilde {\cal M}(H)$ appears in the action, and hence equations of motion, it can be assumed that \begin{equation}
  \tilde {\cal M}(H)=-\star_\eta\tilde {\cal M}(H)\,.
  \label{antiselfM}
\end{equation}
Similarly, it can be assumed that $\tilde {\cal M}$ is also symmetric the  sense that 
\begin{align}\label{sym}
	H_1\wedge {\tilde {\cal M}}(H_2)= H_2\wedge {\tilde {\cal M}}(H_1) \ ,
\end{align}
holds for any two self-dual three-forms $H_1,H_2$. 
We note that in \cite{Sen:2015nph, Sen:2019qit} the following notation is employed
\begin{align}\label{SenM}
{\cal M}^{\mu\nu\lambda;\alpha\beta\gamma} = \frac{4}{ 3!}\varepsilon^{\mu\nu\lambda\rho\sigma\tau}\tilde {\cal M}_{\rho\sigma\tau}^{\alpha\beta\gamma} = -4 \eta^{\mu\rho}\eta^{\nu\sigma}\eta^{\lambda\tau}\tilde {\cal M}_{\rho\sigma\tau}^{\alpha\beta\gamma}\ ,	
\end{align}
where in the last step we used \eqref{antiselfM}.

The role of the last term  in $S_H$ is to change the equations of motion to 
\begin{align}
d\left( \frac12 \star_\eta dB + H\right) & =0\nonumber\\
dB - \tilde {\cal M}(H) &= \star_{\eta}\left(dB -   \tilde {\cal M}(H)\right)	\ ,
\end{align}
which can be recast into
\begin{align}
  dH_{(s)}&=0 \cr
  d\left(H-  \tilde {\cal M}(H)\right)	&=0\ .
\end{align} 

\subsection{A construction for $\tilde {\mathcal M}$}

We would next like to find $\tilde {\cal M}$ such that
\begin{align}\label{Mdef}
\star_g \left(H - {\tilde {\cal M}}(H)\right)= H -  {\tilde {\cal M}}(H)\ ,
\end{align}
for arbitrary $H$, self-dual with respect to $\star_\eta$.
One can then define
\begin{align}
  \label{eq:1}
 H_{(g)} :=  H  -  \tilde {\cal M}(H) \ ,
\end{align}
which satisfies $H_{(g)}  = \star_g H_{(g)} $ by construction  and $d   H_{(g)} =0$ by the equations of motion.   

To achieve \eqref{Mdef}, observe that $\tilde {\cal M}$ is a linear map from self-dual three-forms to anti-self-dual three forms (with respect to $\star_\eta$). However, it is helpful to extend its action to arbitrary 3-forms. Requiring that the symmetry property \eqref{sym} holds for arbitrary 3-forms implies that $\tilde {\cal M}$ should vanish on anti-self-dual three-forms (with respect to $\star_\eta$). This property can be made explicit by re-writing
\begin{align}\label{eq0}
	\tilde {\cal M} \to  \frac14 (1-\star_\eta)\tilde {\cal M}(1+\star_\eta)\ .
\end{align}
 Given that $H = \frac12(1+\star_\eta)H$, the condition (\ref{Mdef}) becomes
\begin{align}\label{eq1}
  	 \frac1{4} (1-\star_g)(1-\star_\eta)\tilde  {\cal M} (1+\star_\eta)
=
\frac12(1-\star_g )(1+\star_\eta) \ ,
\end{align}
and can be viewed as a linear-operator equation acting on arbitrary 3-forms. 

To solve this, we consider a basis of 3-forms  given by
\begin{align}
\omega^A_+\;, \omega_{-A}\qquad\mathrm{for}\qquad A = 1,...,10\ ,
\end{align}
where the subscript $\pm$ indicates their eigenvalue under $\star_\eta$. The number of self-dual and anti-self-dual forms are equal so we have used the same index to label them (but one upstairs and one downstairs).  When acting on this basis we can write ${\tilde {\cal M}}$ in terms of a  matrix $\tilde  {\cal M}^{AB}$:
\begin{align}\label{Mmatrix}
{\tilde  {\cal M}}(\omega_{-A}) =0\ , \qquad {\tilde  {\cal M}}(\omega_+^A) = \tilde  {\cal M}^{AB} \omega_{-B}	\ .
\end{align}
Note that if we choose a basis where
\begin{align}\label{+basis}
\omega^A_+\wedge \omega_{B-} =2\delta^A_B dx^0\wedge...\wedge dx^5\ ,	
\end{align}
then the symmetry condition (\ref{sym}) reduces to $ \tilde {\cal M}^{AB}= \tilde {\cal M}^{BA}$.

Equation (\ref{eq1}) is trivially satisfied when acting on $\omega_{-A}$. However, acting on $\omega^A_+$ gives
\begin{align}
  \tilde  {\cal M}^{AB} (1-\star_g)\omega_{-B} &= (1-\star_g)\omega^A_+\ ,
\end{align}
which  can be  re-arranged to
\begin{align}\label{eq2}
  (1-\star_g) \left(\omega^A_+ -  \tilde {\cal M}^{AB}\omega_{-B} \right)= 0	\ ,
\end{align}
implying that  $\omega^A_+ -  \tilde {{\cal M}}^{AB}\omega_{-B}$ is self-dual with respect to $\star_g$.

Next, we can also  construct a basis $\varphi^A$ of  self-dual three-form solutions with respect to $\star_g$. In particular, at any given point we can  write:
\begin{align}\label{formexpansion}
  \varphi^A  = {\cal N}^A{}_B\omega^B_+ + {\cal K}^{AB}\omega_{-B}\ .
\end{align}
 The condition that $\omega^A_+ -  \tilde {{\cal M}}^{AB}\omega_{-B}$ is self-dual with respect to $\star_g$ implies that we can find a $\Theta^A{}_B$ such that 
\begin{align}
  \omega^A_+ -  \tilde {{\cal M}}^{AB}\omega_{-B} &= \Theta^A{}_{B}\varphi^B \ret
  & = \Theta^A{}_{B}{\cal N}^B{}_C\omega^C_+ + \Theta^A{}_{B}{\cal K}^{BC}\omega_{-C}\ .
\end{align}
Since the $\omega^A_+$ and $\omega_{A-}$ form a basis of three-forms, this implies that
\begin{align}
	 \Theta^A{}_{B}= ({\cal N}^{-1})^A{}_B \ ,
\end{align}
and also results into an expression for $\tilde {\mathcal M}$:
\begin{align}\label{Mdef2}
\tilde  {\cal M}^{AB}	=  - ({\cal N}^{-1})^A{}_C{\cal K}^{CB}\ .
\end{align}
It is important to note that these are all local considerations which are valid at a generic point in spacetime. There could be  global issues as both ${\cal N}^A{}_B$ and ${\cal K}^{AB}$ are only defined locally and  ${\cal N}^A{}_B$ may not be invertible everywhere.
However if at any point ${\cal N}$ is not invertible then there exists a self-dual 3-form with respect to $\star_g$, which is anti-self-dual with respect to $\star_\eta$. However, this is not possible if the spacetime is orientable.

Lastly, let us check that \eqref{Mdef2} is compatible with the symmetry condition $\tilde  {\cal M}^{AB}=\tilde  {\cal M}^{BA}$. To this end we can construct, for any choice of $A$ and $B$, 
\begin{align}
({\cal N}^{-1})^A{}_C\varphi^C &= 	\omega^A_+ - \tilde  {\cal M}^{AC}\omega_{-C}\nonumber\\
({\cal N}^{-1})^B{}_D\varphi^D &= 	\omega^B_+ - \tilde  {\cal M}^{BD}\omega_{-D}\ .
\label{doubling}
\end{align}
These are both self-dual forms with respect to $\star_g$ and therefore their wedge product vanishes:
\begin{align}
0 & = ({\cal N}^{-1})^A{}_C\varphi^C \wedge 	({\cal N}^{-1})^B{}_D\varphi^D \nonumber\\
& =-  \tilde  {\cal M}^{BD}\omega_+^A\wedge \omega_{-D}-  \tilde  {\cal M}^{AC}\omega_{-A}\wedge \omega_+^B\ ,
\end{align}
where we have used the fact that the wedge product of two self-dual or two anti-self-dual forms with respect to $\star_\eta$ also vanishes.
Using the condition (\ref{+basis}) we see that
\begin{align}
0=2 (\tilde {\cal M}^{AB}-\tilde{\cal M}^{BA})dx^0\wedge \ldots\wedge dx^5\ ,	
\end{align}
which ensures that indeed $\tilde{\cal M}^{AB}=\tilde{\cal M}^{BA}$.

It is interesting to observe that, although $H_{(g)}=H -  \tilde {\cal M}(H)$ is self-dual with respect to $\star_g$, it is not typically equal to $\frac12(H + \star_g H)$.  
Rather we find
\begin{align}
	H_{(g)} = \frac12(H + \star_g  H) - \frac12(1 + \star_g)\tilde{\cal M}( H)\ .
\end{align} 
In particular if $H = H_A\omega^A_+$ then (see \eqref{doubling})
\begin{align}\label{HMH}
 H_{(g)}  = ({\cal N}^{-1})^A{}_B H_A\varphi^B\ .
\end{align}
We can introduce a more compact notation as follows: for any (not necessarily self-dual) three-form $\omega$
 we have $\tilde {\cal M}(\tilde {\cal M}(\omega))=0$ so  that   
if we define the map 
\begin{align}\label{mdef}
{\mathfrak m}:\omega \mapsto \omega - \tilde {\cal M}(\omega)	\ ,
\end{align}
then its inverse is
\begin{align}\label{mdef}
{\mathfrak m}^{-1}:\omega \mapsto \omega + \tilde {\cal M}(\omega)	\ .
\end{align}
The map ${\mathfrak m}$ takes $\star_\eta$-self-dual 3-forms to $\star_g$-self-dual 3-forms but acts as the identity on $\star_\eta$-anti-self-dual 3-forms. It does not make all 3-forms $\star_g$-self-dual. 

If $H_{(g)}$ is $\star_g$-self-dual then the map $\mathfrak m$ can be used to write
\begin{align}\label{minv}
H_{(g)} = {\mathfrak m}\left(\frac12(1+\star_\eta) H_{(g)} \right)\ .	
\end{align}
This is due to $\cal{\tilde M}$ being anti-self-dual with respect to $\star_\eta$; see \eqref{antiselfM}. Indeed, if $H_{(g)}$ is $\star_g$-self-dual, there is always an $\star_\eta$-self-dual $H$ such that $H_{(g)}={\mathfrak m}(H)$. We get
\begin{equation}
\frac12(1+\star_\eta) H_{(g)}=\frac12(1+\star_\eta) (H-{\cal{\tilde M}}(H))=H\ ,
\end{equation} and hence \eqref{minv}.
 
\subsection{Introducing Sources}
 
To consider sources $J$, the action we would like to consider is \cite{Sen:2019qit}
\begin{align}
S^J_H= \int \left(\frac12dB\wedge\star_\eta dB -2H\wedge dB +  H\wedge {\tilde {\cal M}}(H)  + 2H\wedge{\tilde {\cal M}}(J) + 2H\wedge J   \right)\ .
\end{align}
As before, $H_{(s)} = \frac12 dB + \frac 12\star_\eta dB +H $ is still a free $\star_\eta$-self-dual form: $H_{(s)}=\star_\eta H_{(s)}$ and, on-shell, $dH_{(s)}=0$. However, 
if we now define 
\begin{align}
	H^J_{(g)} :=\mathfrak{m}(H+J_+)= H +J_+ - \tilde  {\cal M}(H+J_+)  \ ,
	\end{align}
where $J_\pm = \frac12(1\pm \star_\eta) J$,
then the equation of motion becomes
\begin{align}
dH^J_{(g)} = dJ	\ ,
\label{eomnew}
\end{align}
while $H^J_{(g)}=\star_g H^J_{(g)}$ holds by construction.

With the identification $H^J_{(g)} = dA+J$, \eqref{eomnew} is the same equation of motion one would find from the usual action
\begin{align}
S_A=	-\frac12 \int (dA+J)\wedge \star_g (dA+J) +   \int dA\wedge J\ ,
\end{align}
where the self-duality condition $dA+J = \star_g(dA+J)$ must be imposed by hand. One could also add to $S_H$ a term  
\begin{align}
S_J = \int  J\wedge \tilde {\cal M}(J) -\frac12 J\wedge  \star_\eta J\;,
\end{align}
which does not affect the equations of motion but makes the actions $S_H+S_J$ and $S_A$ identical on-shell if we identify $H^J_{(g)} = dA+J$. In this case the complete action can be written as
\begin{align}
S^J_H= \int \left(\frac12 dB\wedge\star_\eta dB -2H\wedge dB +  (H+J_+)\wedge {\tilde {\cal M}}(H+J_+) +2H\wedge J_-  - J_- \wedge J_+	\right)\ .
\end{align}
In addition to the trivial gauge redundancy given by the shift of $B$ by a closed 2-form, this action is also invariant under the following gauge transformation \cite{Sen:2019qit}:
\begin{align}
\delta_\Lambda B &=\Lambda\\
\delta_\Lambda J&= d\Lambda\\
\delta_\Lambda H&=-\Big(\frac{1+\star_\eta}{2}\Big) d\Lambda	\ , 
\end{align}
where $\Lambda$ has to satisfy
\begin{equation}
\int \Lambda\wedge dJ =0\quad.
\end{equation}
Because of this constraint, in general it is not possible to completely gauge away $B$. Notice also that $H_{(s)}$ and $H^J_{(g)}$ are gauge invariant quantities. In particular, the decoupling of $H_{(s)}$ from the physical degrees of freedom does not depend on the gauge choice. One also observes that the above gauge transformations do not commute with diffeomorphisms, even when $\delta_\xi J = 0$.

\subsection{Diffeomorphisms}\label{sec:diffeos}

We now turn to the issue of diffeomorphisms, which are already known to enter the discussion in a novel way from \cite{Sen:2015nph,Sen:2019qit}. Here we will expand on the latter discussion by utilising the construction of $\tilde{\mathcal M}$ from the previous sections.

Let us begin by examining how diffeomorphisms act on the original fields $B$ and $H$. In particular,  consider an infinitesimal coordinate transformation $x^\mu \to x^\mu + \xi^\mu(x)$.  
We will denote the transformation on $B$ by $\delta_\xi B$ and assume that 
\begin{align}
\delta_\xi H = -\Big(\frac{1+\star_\eta}{2}\Big) d\delta_\xi B	\ , 
\end{align}
so that $H_{(s)}$ is invariant: $\delta_\xi H_{(s)}=0$, as one expects from a field that does not gravitate (as we will see later, $H_{(s)}$ completely decouples from the physical degrees of freedom). By neglecting the boundary term $\int d(\delta_\xi B\wedge dB) $ we find
\begin{align}
	\delta_\xi S_H = \int -2(H - \tilde{\cal M}(H)) \wedge d\delta_\xi B +  H\wedge \delta_\xi\tilde {\cal M}(H)\ .
\end{align}  
Note that,  since $\tilde{\cal M}(H)$ and $\delta_\xi\tilde{\cal M}(H)$ are both anti-self-dual with respect to $\star_\eta$,  the second term can be written as $(H - \tilde{\cal M}(H))\wedge \delta_\xi \tilde{\cal M}(H)$ and therefore we can also write this as 
\begin{align}
	\delta_\xi S_H = \int -2H_{(g)}  \wedge d\delta_\xi B +  H_{(g)}\wedge \delta_\xi\tilde {\cal M}(H)\ ,
	\label{deltaS}
\end{align}  
where $H_{(g)}= H - \tilde {\cal M}(H)$.

We now need to ensure that $ H_{(g)}$ remains self-dual with respect to $\star_g$ after the diffeomorphism:
\begin{align}
0=\delta_\xi\left[(1-\star_g)  H_{(g)}\right]=-\delta_\xi\star_g  H_{(g)}	+ (1-\star_g)\delta_\xi H_{(g)}\ .
\label{constraint_delta_Hg}
\end{align}
Note that
\begin{align}
\delta_\xi H_{(g)}  = 
{\mathfrak m}(\delta_\xi H)-\delta_\xi \tilde {\cal M}(H)	\ ,
\end{align}
with  $\delta_\xi H=\star_\eta\delta_\xi H$, so ${\mathfrak m}(\delta_\xi H)$ is $\star_g$-self-dual and on the one hand \eqref{constraint_delta_Hg} simply gives
\begin{align}\label{hhjn}
\delta_\xi\star_g  H_{(g)}=(1-\star_g)\delta_\xi H_{(g)}  = 	-(1-\star_g)\delta_\xi \tilde {\cal M}(H)	 \ .
\end{align}
On the other hand, a direct computation results in 
\begin{align}
\delta_\xi\star_g H_{(g)} &= \nabla_\rho\xi^\rho H_{(g)}	 - \frac12(\nabla_\mu \xi^\rho + \nabla^\rho \xi_\mu)H^{(g)}_{\nu\lambda\rho}dx^\mu\wedge dx^\nu\wedge dx^\lambda\nonumber\;,
\end{align}
where we used that $\delta_\xi g_{\mu\nu} =-2\nabla_{(\mu} \xi_{\nu)}$. Therefore we obtain
\begin{align}\label{deltaM}
	\delta_\xi \tilde {\cal M}(H) &= \frac12 \nabla_\mu\xi^\pi H^{(g)}_{\nu\lambda\pi}dx^\mu\wedge dx^\nu\wedge dx^\lambda  + \Xi - \tilde {\cal M}(\Xi)\ .
\end{align}
Here $\Xi$ is any 3-form which is self-dual with respect to $\star_\eta$ so that the combination 
$\Xi - \tilde {\cal M}(\Xi)$ is self-dual with respect to $\star_g$ and hence does not contribute to (\ref{hhjn}). We will fix $\Xi$ shortly.

To proceed, we observe that
\begin{align}
-2\int H_{(g)}\wedge d (i_\xi H_{(g)}	)  & =  -\frac12\int H_{(g)}\wedge \left(\nabla_\mu \xi^\pi H^{(g)}_{\nu\lambda\pi}dx^\mu\wedge dx^\nu\wedge dx^\lambda\right)\nonumber\\
& = -\int H_{(g)}\wedge \delta_\xi \tilde {\cal M}(H)\ ,
\end{align}
using $H_{(g)}=\star_gH_{(g)}$  and up to a total derivative, with $i_\xi$ the standard inner derivative. Note that once again the $\Xi - \tilde {\cal M}(\Xi)$ term in $\delta_\xi \tilde {\cal M}(H)$ does not contribute here as both it and $H_{(g)}$ are self-dual with respect to $\star_g$ and hence their wedge product vanishes.  Therefore we can define
\begin{align}\label{diffeoB}
\delta_\xi B = 	i_\xi H_{(g)} \ ,
\end{align}
so that $\delta_\xi S=0$, up to a total derivative.

Lastly, we need to fix $\Xi$ to ensure that $	\delta_\xi \tilde {\cal M}(H) $ is anti-self-dual with  respect to $\star_\eta$:
\begin{align}
0&=(1+\star_\eta)	\delta_\xi \big(\tilde {\cal M}(H)\big) 
   \nonumber\\
& = \frac12 (1+\star_\eta)\nabla_\mu\xi^\pi H^{(g)}_{\nu\lambda\pi}dx^\mu\wedge dx^\nu\wedge dx^\lambda +2 \Xi \ ,
\end{align}
where we have used the facts   $(1+\star_\eta) \tilde {\cal M}(\Xi)=0$ and $(1+\star_\eta)  \Xi=2\Xi$. 
Therefore we let 
\begin{align}
\Xi = 	-\frac14 (1+\star_\eta)\nabla_\mu\xi^\pi H^{(g)}_{\nu\lambda\pi}dx^\mu\wedge dx^\nu\wedge dx^\lambda \ ,
\end{align}
and hence, if we introduce the notation 
\begin{align}
\xi( \omega) := \frac1{(p-1)!}\nabla_\mu\xi^\lambda \omega_{\lambda\mu_1...\mu_{p-1}}dx^\mu\wedge dx^{\mu_1}\wedge\ldots\wedge  dx^{\mu_{p-1}}\ , 	
\end{align}
for any $p$-form $\omega$, then 
\begin{align}
	\delta_\xi \tilde {\cal M}(H)	&= \frac12(1-\star_\eta)	\left[  \xi(H_{(g)})   + \tilde {\cal M}\left(   \xi( H_{(g)} ) \right)\right]\nonumber\\ 
	& =\frac12(1-\star_\eta)	\left[ \xi( H) -\xi (\tilde {\cal M}( H))+  \tilde {\cal M}(\xi( H))-\tilde{\cal M}(\xi (\tilde {\cal M}( H))) \right]  \ .
\end{align}
 Note that we also can write this as
\begin{align}
\delta_\xi \tilde {\cal M}(H)  = \Big(\frac{1-\star_\eta}{2}\Big)	{\mathfrak m}^{-1}(\xi({\mathfrak m}(H)))	\ ,
\end{align} 
where the map ${\mathfrak m}(\omega) = \omega- \tilde {\cal M}(\omega)$ was defined in (\ref{mdef}).
This transformation law for $\tilde {\cal M}$ is analogous to that of a connection. In particular, if $\tilde {\cal M}$ vanishes in one frame it need not vanish in another and it is not consistent to set it to zero by fiat in \eqref{SenAction} if one wants to maintain diffeomorphism invariance.
  
We can use the above result to finally determine the transformation properties of $H_{(g)}$. From its definition, we find that
 \begin{align}
 \delta_\xi H_{(g)} &= \delta_\xi H - \tilde {\cal M}(\delta_\xi H)-\delta_\xi \tilde {\cal M}(H)	\nonumber\\
 &= -\xi(H_{(g)})+{\mathfrak m}\left(\frac12(1+\star_\eta)\left(-d(i_\xi H_{(g)})+\xi(H_{(g)})\right)\right)	\ ,
 \end{align}
 but since 
 \begin{align}
 	-d(i_\xi H_{(g)})+\xi(H_{(g)}) 
 	= i_\xi (dH_{(g)} ) -  \xi^\pi \nabla_\pi H_{(g)} \ 
 \end{align}
 we have, on-shell  {\it i.e.} using $dH_{(g)}=0$, that
 \begin{align}\label{Husual}
 \delta_\xi H_{(g)} &= -\xi(H_{(g)}) - m\left(\frac12(1+\star_\eta)  \xi^\pi \nabla_\pi H_{(g)}\right)	 \nonumber\\
 &= -\xi(H_{(g)})- \xi^\pi \nabla_\pi H_{(g)} \nonumber\\
 &= -{\pounds}_\xi H_{(g)}\ ,
 \end{align}
 where we have used the fact that $\star_g \xi^\pi \nabla_\pi H_{(g)}= \xi^\pi \nabla_\pi H_{(g)}$ along with (\ref{minv}), and we denoted the standard Lie derivative with ${\pounds}_\xi$. Thus we recover on shell the usual tensor transformation law  for  $H_{(g)}$ under a diffeomorphism.
 
In the presence of a source $J$ we simply modify \eqref{diffeoB} by considering
\begin{align}
\delta_\xi B = i_\xi H^J_{(g)} - i_\xi J	\ ,
\end{align}
where $H^J_{(g)} =H+J_+ - \tilde {\cal M}(H+J_+)$. Using the usual expression  for the variation of $J$
 \begin{align}
 \delta_\xi J = -\xi(J)- \xi^\pi \nabla_\pi	 J= -{\pounds}_\xi J\ , 
 \end{align}
 we  recover  the standard tensorial variation $\delta_\xi H^J_{(g)} = -\pounds_\xi H^J_{(g)} $ on-shell. 

It is worth emphasising that, although $B$ and $H$ have many properties associated with familiar differential forms, they have non-standard transformations under diffeomorphisms. Therefore, it might be more appropriate to refer to them as ``pseudo-forms''.   

\subsection{Energy-Momentum Tensor}

To further exhibit how the action \eqref{SenAction} reproduces standard results following from diffeo\-morphism-invariant theories, we can use the $\tilde {\cal M}$ term to compute the energy-momentum tensor as the response to the action from a variation of the spacetime metric.\footnote{Here we will set the matter fields to zero as their contribution can be computed by regular means.}

As usual we define 
\begin{align}\label{EMdef}
T_{\mu\nu} &:= -\frac{2}{\sqrt{-g}} \frac{\partial {\cal L}}{\partial g^{\mu\nu}}\nonumber\\
& =  -\frac{2}{\sqrt{-g}} H_AH_B \omega^A_+\wedge \frac{\partial \tilde {\cal M}}{\partial g^{\mu\nu}}(\omega^B_+)\ ,
\end{align} 
where we have expanded $H = H_A\omega^A_+$.
To compute this we note that
\begin{align}
 (1-\star_g)(\omega^B_+- \tilde{\cal M}(\omega^B_+))=0\ , 
\end{align}
which, when varied with respect to the metric $g$, yields
\begin{align}
 (1-\star_g) \delta\tilde{\cal M}(\omega^B_+)  =  -\delta\star_g(\omega^B_+- \tilde{\cal M}(\omega^B_+))\ .
\end{align}
Therefore, for any $\varphi^A = \star_g\varphi^A$,
\begin{align}
2\varphi^A\wedge \delta\tilde{\cal M}(\omega^B_+)  = -\varphi^A\wedge \delta\star_g(\omega^B- \tilde{\cal M}(\omega^B_+))\ ,
\end{align}
and hence from (\ref{HMH}) we find
\begin{align}
2  \delta\tilde{\cal M}^{BC}  \varphi^A_+\wedge\omega_{C-}  =  -({\cal N}^{-1})^B{}_C \varphi^A \wedge \delta\star_g  \varphi^C\ .
\end{align}
On the other hand from (\ref{+basis}) we have
\begin{align}
\varphi^A\wedge\omega_{C-} = {\cal N}^A{}_D\omega^D_+\wedge \omega_{C-}\ ,
\end{align}
and hence 
\begin{align}
  \delta\tilde{\cal M}^{BC}  \omega^D_+\wedge\omega_{C-}  &=-\frac12   ({\cal N}^{-1})^D{}_A({\cal N}^{-1})^B{}_C  \varphi^A \wedge \delta\star_g  \varphi^C\ret
&=-\frac12 \left(\omega^D_+- \tilde {\cal M}(\omega^D_+)\right)\wedge\delta\star_g \left(\omega^B_+- \tilde {\cal M}(\omega^B_+)\right)\ .
\end{align}
Lastly, we contract this with $H_B,H_D$ to find
\begin{align}
T_{\mu\nu} =   \frac{1}{\sqrt{-g}} \left(H- \tilde {\cal M}(H)\right) \wedge \frac{\partial\star_g}{\partial g^{\mu\nu}} \left(H- \tilde {\cal M}(H)\right)\ .	
\end{align}
This has a simple interpretation. We first consider the familiar lagrangian 
\begin{align}
\tilde {\cal L} = -\frac12 \tilde H\wedge\star_g \tilde 	H\ ,
\end{align}
where $\tilde H$ is an arbitrary 3-form and  compute its energy-momentum tensor:
\begin{align}
\tilde T_{\mu\nu}&=\frac{1}{\sqrt{-g}}  \tilde H\wedge \frac{\partial\star_g}{\partial g^{\mu\nu}} \tilde 	H\ret 
&= \frac12 \tilde H_{\mu\lambda\rho}\tilde H_{\nu}{}^{\lambda\rho} - \frac1{12}  g_{\mu\nu}\tilde H_{\lambda\rho\tau}\tilde H^{\lambda\rho\tau}\ .
\end{align}
Then to find our energy-momentum tensor $T_{\mu\nu}$ we set $\tilde H =H - \tilde {\cal M}(H) =   H_{(g)}$ and so
\begin{align}\label{emt}
T_{\mu\nu} =   \frac12 H^{(g)}_{\mu\lambda\rho}g^{\lambda\sigma}g^{\rho\tau} H^{(g)}_{\nu\sigma\tau}\ .
\end{align} 
As usual, we can recover the conservation of the energy-momentum tensor from the translational invariance of the theory. Indeed, consider a constant infinitesimal vector $\xi^\mu$ and re-write \eqref{deltaS} as 
\begin{align}
0=& \int -2dH_{(g)}  \wedge \delta_\xi B +  H\wedge \frac{\partial\tilde {\cal M}(H)}{\partial g^{\mu\nu}}\delta_\xi g^{\mu\nu}\nonumber\\
=&\int -2dH_{(g)}  \wedge \delta_\xi B - \frac{1}{2}T_{\mu\nu}\delta_\xi g^{\mu\nu}\sqrt{-g}d^6x\ ,
\end{align}
where we used \eqref{EMdef}. Thus, by using $\delta_\xi g^{\mu\nu} =2\nabla^{(\mu} \xi^{\nu)}$ and the equation of motion $dH_{(g)}=0$, we recover $\nabla^\mu T_{\mu\nu}=0$.

The above discussion can be straightforwardly extended to include sources by performing the replacement $H_{(g)}= H - \tilde  {\cal M}(H) \; \mapsto \; H^J_{(g)} = H +J_+ - \tilde  {\cal M}(H+J_+)  $.

\subsection{Hamiltonian Formulation}

It will be useful to also express the theory in the hamiltonian formulation; this is the language that was first employed in \cite{Sen:2015nph,Sen:2019qit}. To this end we introduce $i,j =1,2,..,5$. Using self-duality the only independent fields are 
$H_{ijk}$, $B_{ij}$ and  $A_i := B_{0i}$. However, only  $B_{ij}$ has a conjugate momentum:
\begin{align}
\Pi^B_{ij} = 	-\frac12 \left(\partial_0 B_{ij} - 2\partial_{[i}A_{j]}\right) + \frac1{3!}\varepsilon_{ijklm}H_{klm}\ ,
\end{align}
where $\varepsilon_{ijklm}=-\varepsilon_{0ijklm}$. The associated Poisson bracket is 
\begin{align}
\{B_{ij}(\vec x,t),\Pi^B_{kl}(\vec y,t)\}	= \delta_{i[k}\delta_{l]j}\delta(\vec x-\vec y)\ ,
\end{align}
and as a result we find that $A_i$ and $H_{ijk}$  impose the constraints
\begin{align}\label{constraints}
\partial_i\Pi^B_{ij}&=0\nonumber\\	
\frac12\varepsilon_{ijklm}\Pi^B_{lm} & = H_{ijk} - \tilde {\cal M}_{ijk}(H) +\frac32\partial_{[i}B_{jk]}\ . 
\end{align}
Following \cite{Sen:2015nph,Sen:2019qit} we introduce
\begin{align}\label{defPi}
\Pi^\pm_{ij} := \frac12\left(\Pi^B_{ij} \pm \frac14 \varepsilon_{ijklm}\partial_kB_{lm}\right)\ ,
\end{align}
so that the constraints (\ref{constraints})  become 
\begin{align}
\partial_i\Pi^\pm_{ij}&= 0 \nonumber\\
\Pi^-_{ij} & = \frac1{2\cdot 3!}\varepsilon_{ijklm}(H_{klm} - \tilde {\cal M}_{klm}(H) )	\ .
\label{constraints_PiMinus}
\end{align}
In particular, we use the second constraint to determine $H_{ijk}$ as a function of $\Pi^-_{ij}$, $H = H(\Pi^-)$. The dynamical variables are then simply $\Pi^\pm_{ij}$ with Poisson brackets:\footnote{In principle, these should be Dirac brackets but in this particular case they reduce to standard Poisson brackets \cite{Sen:2015nph,Sen:2019qit}.}
\begin{align}\label{PB}
	\{\Pi^\pm_{ij}(\vec{x},t),\Pi^\pm_{kl}(\vec{y},t)\} & = \pm\frac18\varepsilon_{ijklm}\frac{\partial}{\partial x^m}\delta(\vec{x}-\vec{y})\nonumber\\
	\{\Pi^+_{ij}(\vec{x},t),\Pi^-_{kl}(\vec{y},t)\}  & = 0\ .
\end{align} 

Explicit calculation reveals that the hamiltonian density can be written as
\begin{align}\label{Hamiltonianform}
\mathcal H =\Pi_{ij}\partial_0B_{ij}- \mathcal{L}= \mathcal H_+ +\mathcal H_- \ ,
\end{align}
with
\begin{align}
\mathcal H_+&=- \ 2 \Pi^+_{ij}\Pi^+_{ij}-4\Pi^+_{ij} \partial_i A^+_j  \nonumber\\ 
\mathcal H_-&=  2\Pi^-_{ij}\Pi^-_{ij} + \frac13 \varepsilon_{ijklm}\Pi^-_{ij}\tilde {\cal M}_{klm}(H(\Pi^-))+4\Pi^-_{ij} \partial_i A^-_j   \ .
\end{align}

Note that we have introduced two independent constraints to impose $\partial_i\Pi^+_{ij}  =0$ and $\partial_i\Pi^-_{ij}  =0$, rather than the single combined constraint $\partial_i\Pi^B_{ij}=\partial_i( \Pi^+_{ij} + \Pi^-_{ij} )=0$ that is obtained directly from the Legendre transform of the lagrangian. The reason is that in the lagrangian formulation the  constraint  $ \partial_i\Pi^B_{ij}=0$ implies both $\partial_i\Pi^+_{ij}  =0$ and $\partial_i\Pi^-_{ij}  =0$, as the difference vanishes due a Bianchi identity. However, in the hamiltonian formulation there is no Bianchi identity and we need to impose independent constraints to ensure we do not  just impose the less-restrictive constraint $\partial_i( \Pi^+_{ij} + \Pi^-_{ij})=0$. In other words,  $A^+_j +A^-_j $ imposes the constraint $\partial_i\Pi^B_{ij}$ and $A^+_j -A^-_j $ imposes the Bianchi identity on $\partial_i(\Pi^+_{ij} - \Pi^-_{ij})$.
Thus we see that  $\Pi^+_{ij}$ degrees of freedom are unphysical, with the wrong sign for their energy, but are decoupled from the physical $\Pi^-_{ij}$ degrees of freedom.

It is interesting to note that in terms of the original lagrangian variables we have
\begin{align}
\Pi^+_{ij} &= -\frac12 H^{(s)}_{0ij}\nonumber\\
\Pi^-_{ij} & = \frac1{2\cdot 3!}\varepsilon_{ijklm}H^{(g)}_{klm}\nonumber\\
&=\frac12 \sqrt{-g}H_{(g)}^{0ij} \ ,	
\label{Pi_intermsof_Hg}
\end{align}
where indices are raised using $g^{\mu\nu}$. We also observe that
\begin{align}
	2\Pi^-_{ij}\Pi^-_{ij} + \frac13 \varepsilon_{ijklm}\Pi^-_{ij}\tilde {\cal M}_{klm}(H) & =\Pi^-_{ij}\Big(2\Pi^-_{ij}+\frac{1}{3} \varepsilon_{ijklm}\tilde {\cal M}_{klm}(H)\Big)\nonumber\\
	&=\Pi^-_{ij}\Big(\frac{1}{3!}\varepsilon_{ijklm}H_{klm}+\frac{1}{3!} \varepsilon_{ijklm}\tilde {\cal M}_{klm}(H)\Big)\nonumber\\
	&=\Pi^-_{ij}\Big(-H_{0ij}+\tilde {\cal M}_{0ij}(H)\Big)\nonumber\\
	&=-\Pi^-_{ij}H^{(g)}_{0ij}\nonumber\\
	& =  -\frac1{2 }\sqrt{-g} H_{(g)}^{0ij} H^{(g)}_{0ij}\ ,
		\end{align}
where we first used \eqref{constraints_PiMinus}, then the (anti)self-duality properties of $H$ and $\tilde {\cal M}$ with respect to $\star_\eta$, and finally \eqref{Pi_intermsof_Hg}. Thus in terms of the lagrangian variables we see that, after imposing the constraints $\partial_i\Pi^\pm_{ij}=0$,  the hamiltonian can be written as
\begin{align}
\mathcal H =  \left(-\frac12 H^{(s)}_{0ij}H^{(s)}_{0ij} - \sqrt{-g}T^0{}_0\right)\ .
\end{align} 
Here $T^0{}_0 = g^{0\mu}T_{\mu 0}$ where $T_{\mu\nu}$  the energy-momentum tensor found in (\ref{emt}). Therefore, we can construct the hamiltonian by first using familiar geometric techniques to compute $T^0{}_0$ and then re-writing it in terms of
$\Pi^-_{ij}  = \frac1{2 }\sqrt{-g} H_{(g)}^{0ij}$  ({\it i.e.} one is required to solve for $H^{(g)}_{0ij}$ in terms of $H_{(g)}^{0ij}$ and hence $\Pi^-_{ij}$).

As a specific example, let us consider the case of a static-like spacetime with $g_{0i}=0$. In that case we simply find
\begin{align}
	H^{(g)}_{0ij} = g_{00}g_{ik}g_{jl}H_{(g)}^{0kl} = \frac{2}{\sqrt{-g}}g_{00}g_{ik}g_{jl}\Pi^-_{kl}\ ,
\end{align}
and hence
\begin{align}\label{Hamiltonian}
\mathcal H=  -2 \Pi^+_{ij}\Pi^+_{ij}-4  \Pi^-_{ij} \partial_i A^+_j   - \frac{2}{\sqrt{-g}}g_{00}g_{ik}g_{jl}\Pi^-_{ij}\Pi^-_{kl}+4  \Pi^-_{ij} \partial_i A^-_j \ .
\end{align} 

External sources can be included by leaving the definitions of $\Pi^\pm_{ij}$ unchanged but modifying the constraint for  $\Pi^-_{ij}$ to 
\begin{align}
\Pi^-_{ij} &= \frac{1}{2\cdot 3!}\varepsilon_{ijklm}(H^{(g)}_{klm}-J_{klm})\nonumber\\
& = \frac12 \sqrt{-g}(H_{(g)}^{0ij}- J^{0ij})          \ .
\end{align} 
In this case we find, imposing the constraints $\partial_i\Pi^\pm_{ij}=0$ and focussing once again on the case of static spacetimes for which $g_{0i}=0$, 
\begin{align}\label{HJ}
\mathcal H =   -\frac12 H^{(s)}_{0ij}H^{(s)}_{0ij} - &\frac12\sqrt{-g}(H_{(g)}^{J\ 0ij}- J^{0ij})(H^{J}_{(g)0ij}- J_{0ij})  \nonumber\\ 
 &  +\frac1{3!}(\star_\eta J)_{ijk}(J-\star_g J)_{ijk} \ ,	
\end{align}
where the indices are raised using $g^{\mu\nu}$. In terms of the hamiltonian variables $\Pi^\pm_{ij}$ (\ref{Hamiltonian}) remains unchanged but now includes terms quadratic in the sources arising from the last line in (\ref{HJ}).

\subsection{Supersymmetry}

Here we will write the (on-shell) supersymmetric completion of the action \eqref{SenAction}, generalising the results of \cite{Lambert:2019diy} to arbitrary backgrounds.  We will not introduce sources although some cases along these lines were considered in \cite{Lambert:2019diy}. We assume that the six-manifold admits a conformal Killing spinor that satisfies
\begin{align}
\nabla_\mu \epsilon =  \Gamma_\mu\zeta	\ ,
\end{align}
for some $ \zeta =
\frac16\Gamma^\rho\nabla_\rho\epsilon$.\footnote{From this one can derive that $\nabla^2 \epsilon = -\frac{1}{10} R \epsilon$ with $R$ the Ricci curvature. Throughout this section we use the conventions of \cite{Lambert:2010wm}.} The matter fields $X^I$ and $\Psi$
can be covariantly coupled to the non-trivial metric as usual
\begin{align}\label{smat}
S_{mat}= \int  \left(
- \frac12 d X^I\wedge \star_g d X^I  + \frac {i}{2}\bar\Psi\Gamma_\mu dx^\mu \wedge \star_g\nabla \Psi - \frac15 RX^IX^I\right)\ ,
\end{align}
with the action remaining invariant under the extended supersymmetry variations
\begin{align}\label{susy}
\delta X^I &=i\bar\epsilon \Gamma^I\Psi\ret
\delta \Psi &= \Gamma^\mu\Gamma^I\partial_\mu X^I\epsilon  -\frac{2}{3}\Gamma^IX^I\Gamma^\rho\nabla_\rho\epsilon + \delta_H\Psi\ ,
\end{align}
where $\delta_H\Psi$ is yet to be determined. Here all geometric quantities are those associated with a curved spacetime and hence $\{\Gamma_\mu,\Gamma_\nu\} = 2g_{\mu\nu}$. A short calculation shows that the terms in $\delta S_{mat}$ involving $X^I$  cancel out, leaving 
\begin{align}
\delta S_{mat} = 	-\int  i\sqrt{-g}\nabla_\mu\bar\Psi \Gamma^\mu  \delta_H\Psi\ .
\end{align}

Let us now look at $\delta S_H$ and take
\begin{align}\label{susy}
\delta B_{\mu\nu} & = -i\bar\epsilon\Gamma_{ {\mu\nu}}\Psi\ret
\delta H_{\mu\nu\lambda} &  =   \frac{3i }{{2}}  \partial_{[\lambda}(\bar\epsilon\Gamma_{ {\mu\nu}]}\Psi) + \frac{3i }{{2}\cdot 3!}\varepsilon_{\mu\nu\lambda\rho\sigma\tau}\eta^{\rho\alpha} \eta^{\sigma\beta}\eta^{\tau\gamma}\partial_{\gamma}(\bar\epsilon\Gamma_{ {\alpha\beta}}\Psi) \ .
\end{align}
A key observation at this point is that
\begin{align}
  \label{eq:2}
  \delta H = - \Big( \frac{1+ \star_\eta}{2}\Big) d\delta B\ ,
\end{align}
and hence $\delta H_{(s)}=0$, {\it i.e.} we have a reducible representation of   $(2,0)$ supersymmetry where $H_{(s)}$ is a singlet.\footnote{One expects the fact that $H_{(s)}$ is a supersymmetry singlet. It is also a singlet under all diffeomorphisms and supersymmetry acts, roughly speaking, as the square root of a translation.} On the other hand, from $\delta S_H$ we have a non-vanishing contribution from $H\wedge d\delta B$ and an additional term\footnote{Clearly, $\delta \tilde{\mathcal M} =0 $, since $\tilde{\mathcal M} =0$ is a function of the background metric only.} from $\delta (H\wedge \tilde {\cal M}(H)) = -2d\delta B\wedge \tilde {\cal M}(H)$ which combine to give
\begin{align}
\delta S_H & = \int \frac{i}{ 3!}\varepsilon^{\mu\nu\lambda\rho\sigma\tau}\partial_\mu(\bar \Psi\Gamma_{ {\nu\lambda }}\epsilon)\left(H -  \tilde{\cal M}(H)\right)_{ \rho\sigma\tau}\ret
& = \int i \sqrt{-g}\nabla_\mu(\bar \Psi\Gamma_{ {\nu\lambda}}\epsilon) (H-\tilde {\cal M}(H))^{\mu\nu\lambda}\ ,
\end{align}
where we have used the fact that $  H -  \tilde {\cal M}(H)$ is self-dual with respect to $\star_g$ and that the Christoffel terms drop out of a covariant derivative involving anti-symmetrised indices. Everything is now in purely geometric terms.

To continue, we note that if $\nabla_\mu\epsilon =\Gamma_\mu\zeta$ then $\Gamma_{012345}\zeta = -\sqrt{-{\det g}}\; \zeta $, hence $\bar\Psi\Gamma_{\mu\nu\lambda}\zeta$ is self-dual. As a result the $\nabla_\mu\epsilon$ term drops out of $\delta S_H$ and we find 
\begin{align}
\delta S_H &    = \int i\sqrt{-g}\nabla_\mu \bar \Psi\Gamma_{ {\nu\lambda}}\epsilon (H-\tilde {\cal M}(H))^{\mu\nu\lambda}.
\end{align}
It is then easy to check that
\begin{align}
	\delta_H\Psi = \frac{1}{ 3!}\Gamma_{\mu\nu\lambda}(H-\tilde {\cal M}(H))^{\mu\nu\lambda}\epsilon\ ,
\end{align} 
will lead to a supersymmetric action.

In summary, we have that the action $S = S_H+S_{mat}$ is invariant under the on-shell supersymmetry, realised by the transformations
\begin{align}\label{susy2}
\delta X^I & = i\bar\epsilon\Gamma^I\Psi\ret
\delta B_{\mu\nu} &= - i\bar\epsilon\Gamma_{\mu\nu}\Psi\ret
\delta H_{\mu\nu\lambda} &  =   \frac{3i }{{2}}   \bar\epsilon\Gamma_{ {[\mu\nu}}\nabla_{\lambda]}\Psi  + \frac{3i }{{2}\cdot 3!}\varepsilon_{\mu\nu\lambda\rho\sigma\tau}\eta^{\rho\alpha} \eta^{\sigma\beta}\eta^{\tau\gamma} \bar\epsilon\Gamma_{ {\alpha\beta}}\nabla_{\gamma}\Psi\ret
& -\frac{i}{4}\nabla^\rho\bar\epsilon\Gamma_\rho\Gamma_{\mu\nu\lambda}\Psi
 - \frac{i}{4\cdot 3!}\varepsilon_{\mu\nu\lambda\rho\sigma\tau}\eta^{\rho\alpha} \eta^{\sigma\beta}\eta^{\tau\gamma} \nabla^\omega\bar\epsilon\Gamma_\omega\Gamma_{\alpha\beta\gamma}\Psi\ret
 \delta \Psi &= \Gamma^\mu\Gamma^I\partial_\mu X^I\epsilon -\frac{2}{3}\Gamma^IX^I\Gamma^\rho\nabla_\rho\epsilon + \frac{1}{  3!}\Gamma_{\mu\nu\lambda} (H-\tilde {\cal M}(H))^{\mu\nu\lambda}\epsilon\ ,
\end{align}
for any spinor that satisfies $\nabla_\mu \epsilon = \frac16\Gamma_\mu\Gamma^\rho\nabla_\rho\epsilon$ and 
$\Gamma_{012345}\epsilon = \sqrt{-g}\;\epsilon$.

\section{Reductions of the Abelian (2,0) Theory}\label{compactifications}

Having developed this geometric formulation we now turn to its compactification. We will focus on three examples: that of a circle, K3 and a Riemann surface. The first reproduces five-dimensional Maxwell theory, while the second gives a heterotic string transverse to $\mathbb R^5\times \mathbb T^3$. The Riemann-surface reduction leads to the Seiberg--Witten effective action for a four-dimensional ${\cal N}=2$ Yang-Mills gauge theory. The first two cases are consistent with expectations whereas the third gives rise to some new features. In this section we set the fermions to zero for simplicity as we do not expect them to provide any novel physics.

\subsection{Reduction on a Circle}

The simplest case to consider is a six-dimensional manifold with a product metric of the form
\begin{align}
g = \left(\begin{matrix}
\eta_5 & 0 \\ 0&R^2	
\end{matrix}
\right)	\ ,
\label{metric_circle}
\end{align}
where $\eta_5$ is the flat five-dimensional Minkowski metric. From the M-theory point of view, reducing a single M5-brane on a circle produces a D4-brane in type IIA string theory, which in turn is described by five-dimensional supersymmetric Maxwell theory.\footnote{We will explicitly perform the reduction of $S_H$ only; the matter part can be reduced as usual.}

On the one hand we can express the $\omega^A_+$ and $\omega_{-A}$ basis of six-dimensional (anti)self-dual three-forms with respect to $\eta_6$ as
\begin{align}
  \omega^A_+ =  \Omega^A  \wedge dx^5
   + \star_5 \Omega^A \nonumber\\	
  \omega_{-A} =   \Omega^A \wedge dx^5 -\star_5\Omega^A \ ,
  \label{used}
\end{align}
where $\Omega^A$ are a basis of two-forms in five dimensions and $\star_5$ is the Hodge dual constructed from $\eta_5$. On the other hand, a basis of self-dual three-forms with respect to $g$ is 
\begin{align}
\varphi^A &= \Omega^A  \wedge dx^5 
 +\frac{1}{R} \star_5 \Omega^A \nonumber\\	
 &=   \frac{R+1}{2R} \omega^A_+ +  \frac{R-1}{2R}\omega_{-A}\ .
 \label{example_circle_basis}
\end{align}
Then, using the definition \eqref{formexpansion} and the result \eqref{Mdef2}, one can extract that for this case
\begin{equation}
\tilde  {\cal M}^{AB} = -\frac{R-1}{R+1}\delta^{AB}\ ,
\label{example_circle_M}
\end{equation}
which is indeed  symmetric because the forms $\omega_+^A$, $\omega_{-A}$ defined in \eqref{used} satisfy the condition \eqref{+basis}. By expanding the self-dual field $H$ in the above basis, $H = H_A\omega^A_+$, we have
\begin{align}
H_{(g)} &= H - \tilde{\cal M}(H)\nonumber\\
& = H_A\omega^A_+ +\frac{R-1}{R+1}H_A\omega_{-A}\nonumber\\
&= 	\frac{2R}{R+1} H_A\Omega^A\wedge dx^5  + 	\frac{2}{R+1} \star_5H_A\Omega^A\ .
\label{example_cirlce_Hg}
\end{align}
From Eqs.~\eqref{example_circle_basis}-\eqref{example_cirlce_Hg} it is clear that $R$ is dimensionless. This is due to the fact that in this theory we are dealing with the $\star_g$-self-duality condition of $H_{(g)}$ which, for the metric chosen in \eqref{metric_circle}, reads as
\begin{equation}
\begin{split}
{H_{(g)}}_{ijk}=&-\frac{1}{R}\epsilon^{ijkl}{H_{(g)}}_{0l5}\\
{H_{(g)}}_{ij5}=&-\frac{R}{2}\epsilon^{ijmn}{H_{(g)}}_{0mn}\ .
\end{split}
\end{equation}
Thus, to keep the dimensions of $H^{(g)}_{\mu\nu\rho}$ independent of $\mu,\nu,\rho$, we work with a convention where $R$ is dimensionless  and $x^5$ is compact with $x^5\cong x^5 + l$ for some parameter $l$ with dimensions of length. The resulting physical size of the fifth dimension is $lR$.

By implementing the above in Eq.~\eqref{Hamiltonian}, we immediately find
\begin{align}
\mathcal H_- =  \frac{2}{R}\Pi^-_{ab}\Pi^-_{ab}  + 4R\Pi^-_{a5}\Pi^-_{a5}+ 4\Pi^-_{ab}\partial_a A^-_{b}  + 4 \Pi^-_{a5}(\partial_a A^-_5- \partial_5 A^-_a)\ ,
\end{align}
where $a,b=1,2,3,4$. If we truncate to the zero-mode sector along the circle then we can solve the $A^-_{a}$ constraint by writing
\begin{align}
	\Pi^-_{ab} = -\frac{\beta}{4l}\varepsilon_{abcd} \partial_cA_d\ ,
\end{align}  
for some $A_a$ with $\beta$ a unitless normalisation factor that can be fixed {\it ad libitum}. The hamiltonian density reduces to  
\begin{align}\label{Hminusbeta}
\mathcal H_- =  \frac{\beta^2}{8lR}(\partial_aA_b -\partial_bA_a )^2  + 4Rl\Pi^-_{a5}\Pi^-_{a5}+  4l\Pi^-_{a5}\partial_a A^-_5\ ,
\end{align}
while the Poisson bracket (\ref{PB}) becomes\footnote{Note that upon reduction over $x^5$ the five-dimensional delta function
  $\delta(\vec{x}-\vec{y})$ changes to $l^{-1}$ times the four-dimensional delta-function
 $\delta_4(\vec{x}-\vec{y})$.}
\begin{align}
\{A_a(\vec{x},t),\Pi^-_{b5}(\vec{y},t)\}	 = \frac{1}{2\beta}\delta_{ab}  \delta_4(\vec{x}-\vec{y}) \ .
\label{canonical5D}
\end{align}
Thus $A_a$ is canonically conjugate to $\Pi^-_{a5}$ provided that we fix $\beta =1/2$.
We can use this last expression to compute Hamilton's equations 
\begin{align}
\label{eomA}
\partial_0 A_a &= 
                 {8 Rl} \Pi^-_{a5} + 4l\partial_a A^-_{5}\;,\nonumber \\
\partial_0 \Pi_{a5} &= 
                      -\frac{1}{8Rl}\partial_b(\partial_aA_b -\partial_bA_a )\ ,
\end{align}
 which, once combined, yield Maxwell's equations for a gauge potential given by $\{4l A^-_5, A_a\}$. A standard five-dimensional Maxwell lagrangian is then obtained through an inverse Legendre transform; by using \eqref{eomA}, we get
 \begin{align}
\mathcal L_- = & \left(\partial_0 A_a \Pi^-_{a5}- \mathcal H_- \right)\Big|_{\Pi^-_{a5}=\frac{1}{8Rl}(\partial_0A_a-4l\partial_aA^-_{5})}\nonumber\\
 = &	 \frac{1}{32Rl }\left( 2 \Big(\partial_0A_a-4l\partial_aA^-_{5}\Big)^2-(\partial_aA_b -\partial_bA_a )^2 \right) \ , 
\end{align} 
 which scales with $1/R$.
  
 Alternatively, we can also perform the reduction within the lagrangian formalism; this is an instructive exercise which makes even more transparent how this $1/R$ dependence in front of the 5D theory is due to the non-standard coupling of the 6D theory to the metric. By dimensionally reducing the action \eqref{SenAction} on a circle, we get
\begin{align}
S_0= l\int_{\mathbb{R}^{1,4}}& \Big[-\frac{1}{2}d_5B\wedge \star_5 d_5B- \frac1{2l^2} d_5B_5\wedge \star_5 d_5 B_5 \nonumber\\
&+  \frac{2}{l}H_5 \wedge d_5B- \frac{2}{l^2} H_5 \wedge \star_5d_5B_5  -  \frac{2}{l^2}\frac{R-1}{R+1}H_5 \wedge \star_5 H_5 \Big] \ ,
\label{zeroLagrangian0}
\end{align}
where all the fields are to be understood as zero-modes and $B_5, H_5$ stand for $B_5:= lB_{\mu 5}dx^\mu $, $H_5 := \frac{l}{2}H_{\mu\nu 5}dx^\mu\wedge dx^\nu$. The equations of motion yield
\begin{align}
d_5F^{(g)} &=  \frac{1}{R}d_5\star_5 F^{(g)} =0\nonumber\\
d_5F^{(s)} &=   d_5\star_5 F^{(s)} =0\ , 
\end{align}
where $F^{(s)}$ and $F^{(g)}$ are defined by
 \begin{align}
F^{(g)}:=& li_5H^{(g)}  = \frac{2R }{R+1} H_5 \nonumber	\\ 
F^{(s)} :=&  H_5  + \frac12d_5B_5 - \frac{l}{2} \star_5d_5B\ .
\end{align}
Thus we recover two five-dimensional free Maxwell fields.

If one computes the hamiltonian density arising from the compactified lagrangian \eqref{zeroLagrangian0} one finds the same result as compactifying the six-dimensional hamiltonian we considered above (including both $\Pi^+_{ij}$ and $\Pi^-_{ij}$ sectors). Therefore $F^{(s)}$ is unphysical. 

On the other hand, one would like to identify the physical degrees of freedom already at the level of the compactified lagrangian. This is better done in the ``dual frame'', where the 2-form $B$ is dualised to a vector $A^B$. That is, in \eqref{zeroLagrangian0} we introduce a Lagrange multiplier $A^B$, which imposes the Bianchi identity on $Q:=d_5 B$ as follows:
\begin{align}
S_0= l\int_{\mathbb{R}^{1,4}}& \Big[-\frac{1}{2}Q\wedge \star_5 Q+  \frac{2}{l}H_5 \wedge Q \nonumber\\
&- \frac1{2l^2} d_5B_5\wedge \star_5 d_5 B_5- \frac{2}{l^2} H_5 \wedge \star_5d_5B_5  \nonumber\\
&-  \frac{2}{l^2}\frac{R-1}{R+1}H_5 \wedge \star_5 H_5 +\frac{1}{l}Q\wedge d_5 A^B \Big]\ ,
\label{zeroLagrangian_dualised}
\end{align}
so that $A^B$ has mass dimension one. By integrating out $Q$ we get\begin{align}
S_0= \frac{1}{l}\int_{\mathbb{R}^{1,4}}& \Big[- \frac1{2} d_5A^B\wedge \star_5 d_5 A^B- 2 H_5 \wedge \star_5d_5A^B \nonumber\\
&- \frac1{2} d_5B_5\wedge \star_5 d_5 B_5- 2 H_5 \wedge \star_5d_5B_5  \nonumber\\
&-  \frac{4R}{R+1}H_5 \wedge \star_5 H_5 \Big]\ .
\label{zeroLagrangian2}
\end{align}
It is then natural to also integrate out $H_5$, the equations of motion for which impose
\begin{align}
\frac{2R}{R+1}H_5=-\frac{1}{2}d_5\left(A^B+B_5\right)\ .\label{Hinteout}
\end{align}
The action then becomes
\begin{align}
S_0=&\frac{1}{Rl}\frac{1-R}{4}\int_{\mathbb{R}^{1,4}}    d_5A\wedge \star_5d_5 A -\frac{1}{l} \frac{1}{1-R}\int_{\mathbb{R}^{1,4}}   d_5A^B\wedge \star_5 d_5 A^B \ ,
\label{zeroLagrangian2}
\end{align}
where the vector $A$ is defined as
\begin{equation}
A:=B_5+\frac{1+R}{1-R}A^B\quad.
\end{equation}
Here we see two free Maxwell fields with opposite signs for their kinetic terms. When we take $R\to0$, $A$ has the correct sign and its kinetic term scales with $1/R$. In this limit $A=B_5+A^B$ and \eqref{Hinteout} then states that $d_5A$ is nothing but $F^{(g)}$, {\it i.e.} $d_5A=-2F^{(g)}$.

In summary, by performing a circle reduction we have found a five-dimensional lagrangian that scales like $1/R$ rather than $R$; the latter scaling had been previously noted as a challenge for the construction of an action for the M5-brane  \cite{Witten:2009at}. Of course, the discussion here might be somewhat unconvincing as we have a free theory and hence we can rescale the fields by any function of $R$ that we like (recall that $R$ is dimensionless), for example by taking a different choice of $\beta$ in Eqs.~\eqref{Hminusbeta}-\eqref{canonical5D}. However the Poisson bracket  we used  arose from six-dimensions and   its normalisation is fixed. Furthermore the dependence on $l$ is determined by dimensional analysis and only the combination $Rl$ has a physical meaning as the size of the fifth dimension. So there is some hope that this calculation is meaningful.

A more stringent test would be to recover the same $R$ scaling in five-dimensional Super-Yang-Mills by considering the non-abelian action constructed in \cite{Lambert:2019diy}, so we close this subsection by sketching some aspects of the corresponding calculation. The lagrangian of \cite{Lambert:2019diy} employs a covariantly-constant vector field $Y^\mu$ with dimensions of length, first introduced in \cite{Lambert:2010wm}.\footnote{In that construction, $Y^\mu$ takes values in a three-algebra.} For a circle reduction it is natural to fix $Y^5=y$,\footnote{With $y$ some element of the three-algebra.} and hence independent of $x^5$. However, in the cases where $Y$ is not null, it is straightforward to see by looking at the matter terms in the action that the five-dimensional coupling constant will be 
\begin{align}
g^2 = Rl\left(\frac{|\langle y,y\rangle|}{R^2l^2}\right)\ . 	
\end{align}
Thus $g^2$ can be thought of as proportional to $Rl$ but with an arbitrary coefficient given by the dimensionless combination $\langle y,y\rangle/{R^2l^2}$. Comparing with string theory requires us to identify $|\langle y,y\rangle| = (2\pi Rl)^2$. 

\subsection{Reduction on K3}

According to U-duality M-theory on K3 is dual to heterotic string theory on ${\mathbb T}^3$ \cite{Hull:1994ys,Witten:1995ex}. In particular, an  M5-brane wrapped on K3 should give the same dynamics as a heterotic string transverse to ${\mathbb R}^5\times {\mathbb T}^3$. At the worldvolume level this reduction was performed in \cite{Cherkis:1997bx}. We now investigate whether the action \eqref{SenAction} is also consistent with this expectation.

The reduction on K3 can be performed in the hamiltonian formulation. We take the K3 to span the dimensions $x^1,..., x^4$. Since the $\mathcal H_+$ component in \eqref{Hamiltonianform} is independent of any geometric information, it is not clear how to reduce it on K3. However this does not pose a problem since, as we did for the circle reduction, one can simply think of ${\cal H}_+$ as a six-dimensional hamiltonian that decouples from the physical degrees of freedom, and focus on reducing  ${\cal H}_-$. To this end, we recall from \eqref{Pi_intermsof_Hg} that  ($a,b\in\{1,2,3,4\}$) 
\begin{align}
\Pi^-_{ab} &= \frac12 \sqrt{g_{\text K3}} H_{(g)}^{0ab} = \frac14\varepsilon^{abcd}H^{(g)}_{5cd}\nonumber\\
\Pi^-_{5a}& =\frac12 \sqrt{g_{\text K3}} H_{(g)}^{05a} = \frac14\varepsilon^{abcd}H^{(g)}_{bcd} \ .
\end{align}
Next, we make the following ansatz for the Kaluza--Klein reduction of the 3-form fields
\begin{align}
H_{0ab}^{(g)} &=   (-P_A\varphi^A_+ + Q^{A'}\varphi_{A'-})_{ab}\nonumber\\
H_{5ab}^{(g)} &=  (P_A\varphi^A_+ + Q^{A'}\varphi_{A'-})_{ab}\nonumber\\
H^{(g)}_{bcd} & = 0 \ ,
\end{align}
where $\star_{\text K3} \varphi^A_+ = \varphi^A_+$ and $\star_{\text K3} \varphi_{A'-} = -\varphi_{A'-}$, with $\varphi^A_+$, $\varphi_{A'-}$  harmonic 2-forms on K3. In particular, here $A=1,2,...,19$ and $A'=1,2,3$.  Note that for such an ansatz the constraint $\partial_i \Pi^-_{ij}=0$ is automatically satisfied ($\Pi^-_{5a}=0$ and $\partial_b \Pi^-_{ba}=0$ because $\varphi^A_+$ and $\varphi_{A'-}$ are harmonic on K3, hence closed). Note that we have assumed that the usual Kaluza--Klein ansatz can be applied even though, strictly speaking, $H_{(g)}$ is not a differential form. In particular,  we assume that the non-standard transformations arising from diffeomorphisms that we discussed in Sec.~\ref{sec:diffeos} can be absorbed by suitably-modified diffeomorphism transformations of $P_A$ and $Q^{A'}$.

Using this input, one finds
\begin{align}
{\cal H}_- &= -\int_{\text K3} 	\frac12 \sqrt{g_{\text K3}} H_{(g)}^{0ab}H^{(g)}_{0ab}\nonumber\\
 & = -\int_{\text K3} 	\frac14  \varepsilon^{abcd}H^{(g)}_{5cd}H^{(g)}_{0ab}\nonumber\\
 & = \kappa^{AB}	P_AP_B + \kappa_{A'B'}Q^{A'}Q^{B'}\ ,
\end{align}
where we defined
\begin{align}
\kappa^{AB} := \int_{\text K3} \varphi^A_+\wedge \varphi^B_+\ ,\qquad 	\kappa_{A'B'} := -\int_{\text K3} \varphi_{A'-}\wedge \varphi_{B'-}\ ,
\end{align}
which clearly are invertible matrices.

We also need to reduce the Poisson bracket (here $\vec x$ and $\vec y$ denote local coordinates on K3 and $\sigma,\sigma'$ are coordinates in the remaining $x^5$ direction):
\begin{align}
-\frac18\varepsilon_{abcd}\delta_4(\vec x-\vec y)\frac{\partial}{\partial \sigma}\delta(\sigma- \sigma') & = \{\Pi^-_{ab}(\sigma,\vec x, t),\Pi^-_{cd}(\sigma',\vec y, t)\}\\
	&= \frac{1}{16}\varepsilon^{abef}\varepsilon^{cdgh}\{ P_A(\sigma,t)\varphi^{A }_{+ef}(\vec x) +Q^{A'}(\sigma,t)\varphi_{A'-ef}(\vec x),\nonumber\\
	&\qquad \qquad\qquad\ \ \  P_B(\sigma',t)\varphi^{B }_{+ gh}(\vec y)+Q^{A'}(\sigma',t)\varphi_{A'- gh}(\vec y) \}\ ,\nn
	\end{align}
and hence
	\begin{align}
          \varepsilon_{ abcd} \delta_4(\vec x-\vec y) \frac{\partial}{\partial \sigma}\delta(\sigma-\sigma') 	      &= -2\det(g_{\text K3}) \{ P_A(\sigma,t)\varphi^{Aab}_{+}(\vec x) -Q^{A'}(\sigma,t)\varphi^{ab}_{A'-}(\vec x),\nonumber\\
     &\qquad\qquad \ \ \ \ \ \ \ \ \  P_B(\sigma',t)\varphi^{Bcd}_{+ }(\vec y)-Q^{A'}(\sigma',t)\varphi^{cd}_{A'- }(\vec y)\} \ .
	\end{align}
Multiplying by $\varphi^C_{+ab}(\vec x)\varphi^D_{+cd}(\vec y)$, $\varphi^C_{+ab}(\vec x)\varphi_{D'-cd}(\vec y)$ and $\varphi_{C'-ab}(\vec x)\varphi_{D'-cd}(\vec y)$  and integrating over K3$\times$K3 we respectively find
\begin{align}
\{P_A(\sigma,t),P_B(\sigma',t)\} &= -\frac12\kappa^{-1}_{AB}\frac{\partial}{\partial \sigma}\delta(\sigma-\sigma') \nonumber\\
\{P_A(\sigma,t),Q^{B'}(\sigma',t)\} &= 0	\nonumber\\
\{Q^{A'}(\sigma,t),Q^{B'}(\sigma',t)\} &=  \frac12(\kappa^{-1})^{ A'B'}\frac{\partial}{\partial \sigma}\delta(\sigma-\sigma') \ .
\end{align}
This returns the same hamiltonian and Poisson-bracket structure as in  
 \cite{Sen:2019qit} for (anti)-chiral bosons, albeit without having compactified the $x^5$ direction. Moreover, Hamilton's equations give
 \begin{align}
 \frac{\partial P_A}{\partial t} - 	 \frac{\partial P_A}{\partial \sigma} & = 0\nonumber\\
  \frac{\partial Q^{A'}}{\partial t} + 	 \frac{\partial Q^{A'}}{\partial \sigma} & = 0\ ,
 \end{align}
and we have recovered 19 chiral bosons from $P_A$ and 3 anti-chiral bosons from $Q^{A'}$.
 
The above must  be supplemented with the six-dimensional scalar hamiltonian and Poisson bracket
\begin{align}
{\cal H}_{\text{scal}} &=   \frac{ \sqrt{g_{K3}}}{2} \left(\Pi^I\Pi^I +g^{ab}\partial_a X^I\partial_b X^I + \partial_5X^I\partial_5 X^I\right)\cr
	 \{X^I(\sigma, \vec x,t ),\Pi^J(\sigma',\vec y, t)\} &=  \frac{1}{\sqrt{g_{\text K3}}}\delta^{IJ}\delta(\sigma-\sigma') \delta_4(\vec x-\vec y)\ ,
\end{align}
derived from the scalar part of the action \eqref{smat} (the Ricci curvature vanishes in $\mathbb{R}^{1,1}\times\mathrm{K3}$ and this still holds if we compactify the $x^5$ direction). Reducing ${\cal H}_{\text{scal}}$ merely requires taking the scalars and their momenta to be independent of K3, and as a result simply introduces a  factor,  vol(K3),
 \begin{align}
  {\cal H}_{\mathrm{scal}} & = \frac12 \text{vol(K3)}\left( \Pi^I\Pi^I +  \partial_\sigma X^I\partial_\sigma X^I\right)\cr
 \{X^I(\sigma,t),\Pi^J(\sigma',t)\} & =  (\text{vol(K3)})^{-1}\delta^{IJ}\delta(\sigma-\sigma') \ .
\end{align}
If we define
\begin{align}
P^I := \sqrt{\frac{\text{vol(K3)}}{2}}\left( \Pi^I - \partial_\sigma X^I\right)\ ,\qquad 	Q^I := \sqrt{\frac{\text{vol(K3)}}{2}}\left( \Pi^I + \partial_\sigma X^I\right)\ .
\end{align}
we then find
\begin{align}
	{\cal H}_{\mathrm{scal}} = \frac12 P^IP^I+\frac12 Q^IQ^I	\ ,
\end{align}
and
\begin{align}
\{P^I(\sigma,t),P^J(\sigma',t)\} & = -\delta^{IJ}\frac{\partial}{\partial \sigma}\delta(\sigma-\sigma') \nonumber\\
\{P^I(\sigma,t),Q^J(\sigma',t)\} & = 0 \nonumber\\
\{Q^I(\sigma,t),Q^J(\sigma',t)\} & =	\delta^{IJ}\frac{\partial}{\partial \sigma}\delta(\sigma-\sigma')\ .
\end{align}
This leads to 5 chiral bosons $P^I$ and 5 anti-chiral bosons $Q^I$. Similarly, the reduction of the fermionic hamiltonian clearly leads to 8 chiral and 8 anti-chiral fermions in two dimensions.  

Finally, let us impose a flux-quantisation condition of the form
\begin{align}
	\frac{1}{(2\pi )^3}\int_{C_3} H_{(g)} \in \mathbb Z\ ,
\end{align}
over three-cycles $C_3$ in the full six-dimensional theory. For the purposes of this section, it is enough to consider three-cycles of the form $C_3 = S^1\times C_2$, where $S^1$ is the compactified $x^5$ direction with radius $R=1$ (so that $\tilde {\cal M}=0$) and $C_2$ is a two-cycle in K3. The harmonic forms satisfy the quantisation condition 
\begin{align}
	\frac{1}{(2\pi )^2}\int_{C_2} \varphi^A_+\in \mathbb Z\ ,\qquad  \frac{1}{(2\pi )^2}\int_{C_2} \varphi_{A'-} \in \mathbb Z\ ,
\end{align}
 which implies a quantisation condition
\begin{align}
	\frac{1}{ 2\pi  } \int_{S^1} P_A \in \mathbb Z\ , \qquad \frac{1}{ 2\pi  } \int_{S^1} Q^{A'} \in \mathbb Z\ .
\end{align}
This in turn implies  an integral constraint on the zero-modes for $P_A$ and $Q^{A'}$. Thus, if we view $P_A$ and $Q^{A'}$ as arising from chiral bosons $P_A=\partial_\sigma\phi_A$, $Q^{A'}=\partial_\sigma\phi^{A'}$, then $\phi_A$ and $\phi^{A'}$ must be compact with period $2\pi$.

All in all, we find $19+5=24$  chiral bosons (19 of which are compact)  $3+5=8$  anti-chiral bosons (3 of which are compact), $8$ chiral fermions and $8$ anti-chiral fermions    {\it i.e.} the physical degrees of freedom of a  heterotic string transverse to  ${\mathbb R}^5\times {\mathbb T}^3$.

\subsection{Reduction on a Riemann Surface}

It has been known for some time that the dynamics of a single M5-brane on a non-compact Riemann surface leads at low energies to the Seiberg--Witten effective action \cite{Seiberg:1994rs} of a four-dimensional $\mathcal N=2$ gauge theory \cite{Witten:1997sc}. The idea is to wrap the M5-brane worldvolume on a complex curve $\Sigma$, whose embedding into spacetime is specified by some holomorphic function $s(z)$. Such a curve is subjected to boundary conditions whose interpretation at infinity is that of intersecting  M5-branes. Reducing to type IIA string theory leads to a picture of parallel D4-branes suspended between NS5-branes whose dynamics is given by an ${\cal N}=2$ Yang-Mills gauge theory. One then finds that $s(z)$ depends on various moduli of the Riemann surface $u_\alpha$, $\alpha = 1, ..., N-1$. To compute the four-dimensional effective action from the M5-brane one is not interested in all of its dynamics, rather just those of its zero-modes: the moduli $u_\alpha$ and their superpartners.

This framework was used to reproduce the scalar sector of the resultant four-dimensional effective action in \cite{Howe:1997hxz}, where a simple kinetic term for the single M5-brane theory can be easily  written down. To find the dynamics of the vector fields without an action is more involved. Without scalars, and for a flat torus, the calculation appeared in \cite{Verlinde:1995mz}. For the case of a single M5-brane on a generic Riemann surface  the calculation was done in \cite{Lambert:1997dm} using the equations of motion.  This led to interesting integrals over non-holomorphic functions whose evaluation is nevertheless a holomorphic function of the moduli. 

But now that we have a proposed action for the self-dual tensors in six-dimensions, this setup  provides a natural and non-trivial  testing ground for its interpretation as capturing the low-energy dynamics of single M5-brane. The reduction of the action \eqref{SenAction} over a rigid compact torus was already performed in \cite{Sen:2019qit} and shows the correct $SL(2,{\mathbb Z})$ invariance expected from large diffeomorphisms. Here we will concern ourselves with the case of generic, non-compact Riemann surfaces.

We therefore want to consider an M5-brane where two of its directions ($x^4$ and $x^5$ combined into the complex coordinate $z= x^4+ix^5$), are embedded into spacetime by means of the function $s=X^6+iX^{10}$. Here $X^{10}$ denotes the M-theory direction and is compact. We label the remaining worldvolume coordinates by $x^m$, $m=0,1,2,3$. The embedding of the M5-brane is defined by $X^m = x^m$, $X^7=X^8=X^9=0$ and in particular is such that $s(z)$ is a holomorphic function \cite{Howe:1997hxz}.\footnote{At this stage we neglect terms with $\partial_m s\ne 0$ as these will result into higher-order derivative terms in the Seiberg--Witten effective action.} The induced metric on the M5 is given by
\begin{align}
g = \left(\begin{matrix}
\eta_4 & 0 & 0 \\ 0& 0&(1+ \partial_z s\bar \partial_{\bar z} \bar s)/2	\\
0&( 1+\partial_z s \partial_{\bar z} \bar s	)/2&0
\end{matrix}
\right)	\ .
\end{align}
Here the coordinates are $0,1,2,3,z,\bar z$ so that
\begin{align}
\eta = \left(\begin{matrix}
\eta_4 & 0 & 0 \\ 0& 0&1/2  \\
0& 1/2&0
\end{matrix}
\right)	\ .
\end{align}

In the usual fashion, the zero-mode dynamics can be determined by working in the Manton approximation \cite{Manton:1981mp}: the moduli---and consequently $s$---are promoted to functions of the remaining four coordinates $x^m$, $m=0,1,2,3$ that are slowly varying so that \cite{Howe:1997hxz}
\begin{align}
\partial_m  s = \sum_\alpha\frac{\partial s}{\partial u_\alpha}\partial_m u_\alpha\ . 	
\end{align}
From this one defines the Seiberg--Witten differential $\lambda_{SW} = s(z)d z$ \cite{Mikhailov:1997jv} and  the holomorphic   1-forms
\begin{align}
	\lambda_\alpha =  \frac{\partial s}{\partial u_\alpha}dz\ . 
\end{align}
Following \cite{Seiberg:1994rs} one identifies the low-energy scalar fields as
\begin{align}
a_\alpha  = \oint_{A_\alpha} sdz\ ,
\end{align}
where $A_\alpha, B^\alpha$ are a basis of cycles of $\Sigma$ with intersection matrix
\begin{align}
	A_\alpha \cap B^\beta = -B^\beta  \cap A_\alpha  = \delta_\alpha^\beta\ .
\end{align}
One also defines the dual variables $a^D_{\alpha}$ as
\begin{align}
a^D_\alpha  = \oint_{B^\alpha} sdz\ .
\end{align}
The periods of the holomorphic 1-forms are then
\begin{align}
\oint_{A_\gamma} \lambda_\alpha  = \frac{\partial a_\gamma}{\partial u_\alpha}\ , \qquad
 \oint_{B_\gamma} \lambda_\alpha  = \frac{\partial a^D_\gamma}{\partial u_\alpha} \ ,
\end{align}
while the period matrix can be expressed as
\begin{align}
\tau_{\alpha\beta}	 & = \frac{\partial a^D_\alpha}{\partial a_\beta} = \tau_{\beta\alpha}\ .
\end{align}
It is useful to note that
\begin{align}
\int_\Sigma  \lambda_\alpha\wedge \bar \lambda_\beta & = \sum_\gamma	 \left(\oint_{A_\gamma}\lambda_\alpha\oint_{B_\gamma}  \bar \lambda_\beta- \oint_{B_\gamma}\lambda_\alpha\oint_{A_\gamma} \bar \lambda_\beta\right)\nonumber\\
& = \sum_\gamma	 \left(\frac{\partial a_\gamma}{\partial u_\alpha}\frac{\partial \bar a^D_\gamma}{\partial \bar u_\beta}-\frac{\partial  a^D_\gamma}{\partial   u_\alpha}\frac{\partial \bar a_\gamma}{\partial \bar u_\beta}\right)\nonumber\\
& =\sum_\gamma	 \frac{\partial a_\gamma}{\partial u_\alpha}\frac{\partial \bar a_\delta}{\partial \bar u_\beta}(\bar \tau_{\gamma\delta}-  \tau_{\gamma\delta}) \ .
\end{align}
We can also consider the holomorphic 1-forms
\begin{align}
\vartheta_\alpha =	\frac{\partial s}{\partial a_\alpha} dz = \sum_\beta \frac{\partial u_\beta}{\partial a_\alpha}\lambda_\beta\ ,
\end{align}
which are  normalised to have unit period over the $A$-cycles:
\begin{align}
\oint_{A_\gamma}\vartheta_\alpha =\delta_{\alpha}^{\gamma}
\end{align}
and hence
\begin{align}
\oint_{B^\gamma}\vartheta_\alpha =\tau_{\alpha\gamma}\ ,\qquad\int_\Sigma \vartheta_\alpha\wedge \bar\vartheta_\beta = \bar \tau_{\alpha\beta}-\tau_{\alpha\beta}\ .
\end{align}
 
This machinery can be applied to the scalar part of the action \eqref{smat}. One straightforwardly finds \cite{Howe:1997hxz}:
\begin{align}\label{Sscal}
S_{scal} &= -\frac{1}{2}\int d^4x d^2z \partial_ms\partial^m \bar s \nonumber\\
	&= -\frac{1}{2}\sum_{\alpha,\beta}\int d^4x d^2z\ \frac{\partial s}{\partial u_\alpha}\frac{\partial \bar s}{\partial \bar u_\beta}\partial_m u_\alpha\partial^m \bar u_\beta \nonumber\\
	& = -\frac{i}{4}\sum_{\alpha,\beta}\int d^4x \partial_m u_\alpha\partial^m \bar u_\beta \int_\Sigma \lambda_\alpha\wedge \bar \lambda_\beta\nonumber\\
	 & = -\frac{i}{4}\sum_{\alpha,\beta}\int d^4x(\bar \tau_{\alpha\beta}-\tau_{\alpha\beta} ) \partial_m a_\alpha\partial^m \bar a_\beta  \nonumber\\
	 & = -\frac12 \sum_{\alpha,\beta}\int d^4x{\rm Im}\left(\tau_{\alpha\beta} \partial_m a_\alpha\partial^m \bar a_\beta\right)  \ ,
\end{align}
which is precisely the scalar part of the  Seiberg--Witten effective action.  

However, our main goal is to use the action \eqref{SenAction} to reproduce the gauge-field part of the four-dimensional effective action. To proceed, note that if $H$ is of the special form $H = {\cal F}\wedge dz$ or  $H= \bar {\cal F}\wedge d\bar z$ then   one finds $\star_\eta H = H$ provided that $\star_4{\cal F} = -i{\cal F}$. The remaining (anti)self-dual forms with respect to $\eta$ can be expressed in terms of the basis
\begin{align}
\omega_+ &= h + \frac{i}{2}\star_4 h\wedge dz\wedge d\bar z\nonumber\\
\omega_- &=   h - \frac{i}{2} \star_4 h\wedge dz\wedge d\bar z\ ,
\end{align}
where $h = \frac {1}{3!}h_{mnl}dx^m\wedge dx^n\wedge dx^l$.
Therefore in general we have
\begin{align}\label{Hriemann}
H = {\cal F}\wedge dz + \bar {\cal F}\wedge d\bar z	 + h+ \frac{i}{2}\star_4 h\wedge dz\wedge d\bar z\ ,
\end{align}
with ${\cal F} = i\star_4 {\cal F}$, for which $H$ is real and satisfies $\star_\eta H  = H$.

For completeness, let us also determine $H_{(g)}$. When $H = {\cal F}\wedge dz$ or  $H= \bar {\cal F}\wedge d\bar z$ one has that $\star_g H =H$ and thus  $\tilde {\cal M}(H_{mn z} d x^m \wedge dx^n \wedge d z)=\tilde {\cal M}(H_{mn \bar z}  d x^m \wedge dx^n \wedge d \bar z) =0$, whereas the remaining $\star_g$-self-dual forms can be expressed in terms of the basis
\begin{align}
\varphi &= h + i\frac{1+\partial_z s \partial_{\bar z} \bar s}{2}\star_4 h\wedge dz\wedge d\bar z\nonumber\\
&= \frac{ 2+\partial_z s\partial_{\bar z} \bar s	}{2} \omega_+ - \frac{\partial_z s \partial_{\bar z} \bar s	}{2}\omega_-\;,
\end{align}
from which using \eqref{Mmatrix}, \eqref{formexpansion} and \eqref{Mdef2} we obtain
\begin{align}
\tilde {\cal M}\left(h + \frac{i}{2}(\star_4 h)\wedge dz\wedge d\bar z\right) = \frac{\partial_z s \partial_{\bar z} \bar s}{2+\partial_z s\partial_{\bar z} \bar s	}\left(	h - \frac{i}{2} (\star_4 h)\wedge dz\wedge d\bar z\right)\;.
\end{align}
Finally, from \eqref{HMH} 
\begin{align}
H_{(g)} = 	{\cal F}\wedge dz + \bar {\cal F}\wedge d\bar z	 + \frac{2}{2+\partial_z s \partial_{\bar z} \bar s}h+i \frac{1+\partial_z s\partial_{\bar z} \bar s	}{2+\partial_z s \partial_{\bar z} \bar s}\star_4 h\wedge dz\wedge d\bar z\ .
\end{align}

To arrive at the desired four-dimensional effective action including gauge fields, one needs to consider a suitable ansatz for $H$ and $B$ by truncating to the lowest Kaluza--Klein modes; this corresponds to restricting to harmonic 1-forms on $\Sigma$.\footnote{Since $\Sigma$ is non-compact the zero-form and two-form harmonic forms have divergent integrals and hence do not lead to low-energy modes. Thus  $\tilde {\cal M}$ does not play a role here.} We pick the following normalisation:
\begin{align}\label{Hriemann}
H 
& =  \sum_\alpha{\cal F}_\alpha\wedge \vartheta_\alpha	+ \sum_\alpha\bar {\cal F}_\alpha\wedge \bar\vartheta_\alpha	 \ ,
\end{align}
where ${\cal F}_\alpha = i\star_4{\cal F}_\alpha$, while for $B$ we initially set
\begin{align}
B & =   \sum_\alpha A_\alpha\wedge  \vartheta_\alpha+  \sum_\alpha \bar A_\alpha\wedge \bar  \vartheta_\alpha\ ,
\end{align}
where $A_\alpha= A_{\alpha m}dx^m$ are four-dimensional 1-forms.

At this stage recall that the action \eqref{SenAction} has a gauge symmetry $B\to B+d\Lambda$, where $\Lambda$ is an arbitrary 1-form. This is expected to descend to a 0-form gauge symmetry for $A_\alpha$: $A_\alpha \to A_\alpha +d_4 \lambda_\alpha$. However, since the $\vartheta_\alpha$ are dynamical,  under such a transformation
\begin{align}
B \to B +d\left(\sum_\alpha 	\lambda_\alpha\vartheta_\alpha + \sum_\alpha 	\bar \lambda_\alpha\bar \vartheta_\alpha\right) -\sum_{\alpha,\beta} 	\lambda_\alpha d_4a_\beta\wedge \frac{\partial\vartheta_\alpha}{\partial a_\beta} - \sum_{\alpha,\beta} 	\bar \lambda_\alpha d_4\bar a_\beta\wedge \frac{\partial\bar \vartheta_\alpha}{\partial \bar a_\beta} \ .
\end{align}
To compensate for this we introduce  four-dimensional Stueckelberg-like scalar fields $c_\alpha,\bar c_\alpha$ and expand
\begin{align}
B  
& =   \sum_\alpha\left( A_\alpha\wedge  \vartheta_\alpha - c_\alpha d_4a_\beta\wedge \frac{\partial\vartheta_\alpha}{\partial a_\beta}\right)+  \sum_\alpha \left(\bar A_\alpha\wedge \bar  \vartheta_\alpha  -\bar c_\alpha d_4\bar a_\beta\wedge \frac{\partial\bar \vartheta_\alpha}{\partial \bar a_\beta}\right)\ , 
\end{align}
so that under the combined gauge transformation $A_\alpha \to A_\alpha + d_4\lambda_\alpha$, $c_\alpha\to c_\alpha-\lambda_\alpha$ we recover a one-form gauge transformation
\begin{align}
B \to B +d\left(\sum_\alpha 	\lambda_\alpha\vartheta_\alpha + \sum_\alpha 	\bar \lambda_\alpha\bar \vartheta_\alpha\right)   \ .
\end{align}
With this in hand, we compute 
\begin{align}\label{Briemann}
dB & = 	\sum_\alpha d_4A_\alpha\wedge \vartheta_\alpha+ \sum_\alpha d_4\bar A_\alpha\wedge \bar\vartheta_\alpha \\ &-\sum_\alpha (A_\alpha+d_4c_\alpha)\wedge d_4a_\beta  \wedge \frac{\partial \vartheta_\alpha}{\partial{a_\beta}}  - \sum_{\alpha,\beta} (\bar A_\alpha+d_4\bar c_\alpha)\wedge d_4\bar a_\beta\wedge \frac{\partial \bar \vartheta_\alpha}{\partial{\bar a_\beta}} \nonumber\\
\star_\eta dB & = \sum_\alpha i\star_4 d_4A_\alpha\wedge \vartheta_\alpha-  \sum_\alpha i\star_4 d_4\bar A_\alpha\wedge \bar\vartheta_\alpha\nonumber\\ &-  \sum_\alpha i\star_4((A_\alpha+d_4c_\alpha)\wedge d_4a_\beta)\wedge \frac{\partial \vartheta_\alpha}{\partial{a_\beta}}  +  \sum_{\alpha, \beta} i\star_4(  (\bar A_\alpha+d_4\bar c_\alpha)\wedge d_4\bar a_\beta )\wedge \frac{\partial \bar \vartheta_\alpha}{\partial{\bar a_\beta}}\;,\nn
\end{align}
 where $d_4$ denotes the exterior derivative along $x^m$. To continue, observe that
 \begin{align}
 	\int_\Sigma \vartheta_\alpha \wedge \frac{\partial \bar \vartheta_\beta}{\partial \bar a_\gamma} &= \frac{\partial  }{\partial \bar a_\gamma}\int_\Sigma \vartheta_\alpha \wedge  \bar \vartheta_\beta =\frac {\partial \bar \tau_{\alpha\beta}}{\partial \bar a_\gamma}\nonumber\\
 	\int_\Sigma \frac{\partial\vartheta_\alpha}{\partial a_\gamma }\wedge  \bar \vartheta_\beta &= \frac{\partial  }{\partial a_\gamma}\int_\Sigma \vartheta_\alpha \wedge  \bar \vartheta_\beta =-\frac {\partial   \tau_{\alpha\beta}}{\partial  a_\gamma}\ ,
 \end{align}
  and 
   \begin{align}
 	\int_\Sigma \frac{\partial\vartheta_\alpha}{\partial a_\gamma} \wedge \frac{\partial \bar \vartheta_\beta}{\partial \bar a_\delta}   = \frac{\partial  }{\partial \bar a_\delta}\int_\Sigma  \frac{\partial \vartheta_\alpha}{\partial a_\gamma} \wedge  \bar \vartheta_\beta  = -\frac{\partial  }{\partial \bar a_\delta}\left(\frac {\partial   \tau_{\alpha\beta}}{\partial  a_\gamma}\right)=0 \ . 
 	\end{align}
        Substituting \eqref{Hriemann} and \eqref{Briemann} into \eqref{SenAction} we find\footnote{We remind the reader that due to our Kaluza--Klein ansatz $\tilde{\mathcal M}$ does not enter this calculation.}
 \begin{align}\label{SHriemann}
 S_H   &=   \int \Big( -(\tau-\bar\tau)_{\alpha\beta}  \left(
     - d_4A_\alpha\wedge i\star_4 d\bar A_\beta  -2{\cal F}_\alpha\wedge  d_4\bar A_\beta    
  +2\bar{\cal F}_\alpha\wedge  d_4A_\beta  \right) \\
  &\hskip2.9cm +  \frac{\partial\tau_{\alpha\beta}}{\partial a_\gamma}(-i
  \star_4 d_4\bar A_\alpha\wedge   ( A_\beta+d_4  c_\beta) \wedge d_4  a_\gamma+
  2\bar {\cal F}_\alpha\wedge    ( A_\beta+d_4  c_\beta)\wedge d_4  a_\gamma)   \nonumber\\
 &\hskip2.9cm + \frac{\partial\bar \tau_{\alpha\beta}}{\partial \bar a_\gamma} (i
  \star_4 d_4A_\alpha\wedge  ( \bar A_\beta+d_4  \bar c_\beta)\wedge d_4\bar a_\gamma+2{\cal F}_\alpha\wedge  ( \bar A_\beta+d_4  \bar c_\beta)\wedge d_4\bar a_\gamma)\Big) \nonumber\\
  &=   \int \Big( (\tau-\bar\tau)_{\alpha\beta}  \left(
      d_4A_\alpha\wedge i\star_4 d\bar A_\beta  +2{\cal F}_\alpha\wedge  d_4\bar A_\beta    
  -2\bar{\cal F}_\alpha\wedge  d_4A_\beta  \right) \nonumber\\
  &\hskip2.9cm +  (-i
  \star_4 d_4\bar A_\alpha\wedge  ( A_\beta+d_4  c_\beta) \wedge d_4  \tau_{\alpha\beta} +
  2\bar {\cal F}_\alpha\wedge  ( A_\beta+d_4  c_\beta)\wedge d_4  \tau_{\alpha\beta})   \nonumber\\
 &\hskip2.9cm + (i
  \star_4 d_4A_\alpha\wedge  ( \bar A_\beta+d_4  \bar c_\beta)\wedge d_4 \bar\tau_{\alpha\beta}+2{\cal F}_\alpha\wedge  ( \bar A_\beta+d_4  \bar c_\beta)\wedge d_4\bar \tau_{\alpha\beta})\Big)   \ .\nn
  \end{align}
 It is helpful to introduce the  two-form
 \begin{align}\label{Fs}
 {\cal F}^{(s)}_\alpha :=   {\cal F}_\alpha + \frac12 d_4 A_\alpha  + \frac{i}{2}\star_4d_4 A_\alpha \ ,
 \end{align}
 and combine the pieces \eqref{Sscal} and \eqref{SHriemann} to rewrite the action as
 \begin{align}\label{Sfull}
 S    &=  S_{scal}+S_H\nonumber\\
 & =  \int \Big(  -\frac14(\tau-\bar\tau)_{\alpha\beta}  d_4a_\alpha\wedge i\star_4 d_4\bar a_\beta-(\tau-\bar\tau)_{\alpha\beta}  d_4A_\alpha\wedge i\star_4 d_4\bar A_\beta+(\tau+\bar\tau)_{\alpha\beta}  d_4A_\alpha\wedge  d_4\bar A_\beta\nonumber\\
 &\qquad  +2{\cal F}^{(s)}_\alpha \wedge \big((\tau-\bar\tau)_{\alpha\beta}  d_4\bar A_\beta-d_4\bar \tau_{\alpha\beta}\wedge( \bar A_\beta+d_4  \bar c_\beta)\big)\nonumber\\ &\qquad  -2\bar {\cal F}^{(s)}_\alpha\wedge \big((\tau-\bar\tau)_{\alpha\beta}  d_4  A_\beta+d_4\tau_{\alpha\beta}\wedge (   A_\beta+d_4  c_\beta)\big)\Big)\;.
  \end{align}
The first line agrees with the Seiberg--Witten effective action \cite{Seiberg:1994rs} but for two sets of $U(1)$ gauge fields, corresponding to the real and imaginary parts of $A_\alpha$. However, the $\bar {\cal F}^{(s)}_\alpha$ equation imposes the constraint
\begin{align}\label{ceq}
  (\tau -\bar \tau)_{\alpha\beta}  d_4 A_\beta+ d_4 \tau_{\alpha\beta}\wedge( A_\beta + d_4  c_\beta )  =   i\star_4\big(	 (\tau -\bar \tau)_{\alpha\beta}  d_4 A_\beta+ d_4 \tau_{\alpha\beta}\wedge( A_\beta + d_4  c_\beta ) \big)\ .
\end{align}
This implies that the real and imaginary parts of $A_\alpha$ are related by electric-magnetic duality, {\it e.g.} if $d_4\tau_{\alpha\beta}=0$ then this reduces to
\begin{align}\label{SDc}
{\rm Im} (d_4 A_\alpha ) = \star_4 {\rm Re}(d_4  A_\alpha	)\ .
\end{align} 
More generally the constraint (\ref{ceq}) is harder to disentangle. It is worth observing that  $A_\alpha + d_4c_\alpha$ are gauge invariant 1-forms which could provide a restriction on the types of fields that can arise.

Next, we observe that the Stueckelberg fields  impose the equation of motion
\begin{align}\label{stueckel}
	d_4{\cal F}^{(s)}_\alpha\wedge d_4\bar \tau_{\alpha\beta}=0\ ,
\end{align}
which generically implies that  $d_4{\cal F}^{(s)}_\alpha =0$.
Thus the ${\cal F}^{(s)}_\alpha$ decouple in the sense that their equations of motion do  not depend on the other fields.

One also finds extra contributions to the scalar and vector equations of motion arising from ${\cal F}^{(s)}$. Assuming   $d_4{\cal F}^{(s)}_\alpha=0$ we find
\begin{align}
0=&
 (\tau-\bar\tau)_{\alpha\beta}d_4i\star_4 d_4 a_\beta +\frac{\partial\tau_{ \alpha\beta }}{\partial a_\gamma} d_4   a_\gamma\wedge i\star_4 d_4 a_{\beta}   \nonumber\\ &+2\frac{\partial\bar\tau_{\beta\gamma}}{\partial \bar a_\alpha}(d_4A_\beta+i\star_4d_4A_\beta)\wedge (d_4\bar A_\gamma+i\star_4d_4\bar A_\gamma) 
+ 4\frac{\partial\bar\tau_{\beta\gamma}}{\partial \bar a_\alpha}  \bar {\cal F}^{(s)}_\beta\wedge (d_4A_\gamma -i\star_4 dA_\gamma) \nonumber\\
0 = & d_4\left(i\star_4 (\tau-\bar\tau)_{\alpha\beta}dA_\beta-(\tau+\bar\tau)_{\alpha\beta}dA_\beta\right) - 2d_4\tau_{\alpha\beta}\wedge {\cal F}^{(s)}_\beta \ .
\end{align}

One recovers the standard Seiberg--Witten equations \cite{Seiberg:1994rs} in the special case of ${\cal F}^{(s)}_\alpha=0$. More generally, ${\cal F}^{(s)}_\alpha$ acts as an non-dynamical background  electromagnetic field. Its effects can  also be implemented by replacing the last two lines of (\ref{Sfull}) by 
\begin{align}
{\cal L}_{background} = 2\tau_{\alpha\beta}{\cal F}^{(s)}_\alpha\wedge d_4\bar A_\beta	+2\bar \tau_{\alpha\beta}\bar{\cal F}^{(s)}_\alpha\wedge d_4 A_\beta	\ ,
\end{align}
and imposing the self-duality constraint (\ref{ceq}) by hand. 

Lastly, let  us comment on the fact that the equations of motion also admit a sector where $d_4\bar \tau_{\alpha\beta}=0$. On the one hand, for generic $\bar\tau_{\alpha\beta}$ the dynamical constraint on ${\cal F}^{(s)}$ from \eqref{stueckel} freezes out the scalars, and hence also the vectors. On the other, if $\tau_{\alpha\beta}$ is constant then we recover a free Seiberg--Witten theory for two gauge fields related by (\ref{SDc}). Mixed solutions where both $d{\cal F}^{(s)}_\alpha$ and $d_4\bar\tau_{\alpha\beta}$ are non-zero do not seem very likely unless $\tau_{\alpha\beta}$ has some reduced dependence on the moduli. 

\section{Conclusions}\label{sec:conclusions}

In this paper we studied the six-dimensional action put forward in \cite{Sen:2019qit}---and its (2,0) supersymmetric completion \cite{Lambert:2019diy}---clarifying many of its unconventional features. This formulation aims to encode the dynamics of a chiral 2-form in 6D into a ``2-form'' $B$ and an $\star_\eta$-self-dual ``3-form'' $H$. Although all of our analysis is performed for chiral 2-forms in six-dimensions we hope that the techniques we developed can be readily applied to other dimensions.

We elucidated on the coupling of these fields to arbitrary geometries, and in the course of doing so provided a construction of the interaction term $\tilde{\mathcal{M}}$ that goes beyond the perturbative approach of \cite{Sen:2019qit}. Moreover, we wrote down how the original fields $B, H$---which are not conventional differential forms and we dubbed ``pseudo-forms''---can be combined into the unphysical $H_{(s)}$ and the physical $H_{(g)}$ fields; the $H_{(s)}$ is a singlet while $H_{(g)}$ has (on-shell) standard transformation properties under diffeomorphisms. We also clarified some aspects of the hamiltonian analysis. First, we showed that $H_{(s)}$ and $H_{(g)}$ correspond precisely to the $\Pi^{\pm}$ variables introduced in \cite{Sen:2019qit}. The fact that $\Pi^{+}$ describes a non-unitary decoupled sector of the theory is consistent with the fact that $H_{(s)}$ is self-dual with respect to $\eta$. Indeed, $H_{(s)}$ inherits the non-unitarity of $B$ so it must completely  decouple from the physics including gravity. Instead, $H_{(g)}$ carries the physical degrees of freedom and is self-dual with respect to the actual physical metric. Second, we gave a formulation of the hamiltonian in terms of $H_{(s)}$ and the energy-momentum tensor $T^0{}_0$ of $H_{(g)}$. We therefore showed that it is possible to construct the physical hamiltonian by first using familiar geometric techniques to compute $T^0{}_0$ and then re-expressing $H_{(g)}$ in terms of $\Pi^-$; this leads to particularly simple expressions for static backgrounds ({\it i.e.} $g_{0i}=0$).

We then dimensionally reduced the proposed (2,0) action on three backgrounds: a circle, K3 and a Riemann surface. We performed these reductions by implementing the usual Kaluza--Klein ansatz, that is assuming that the only surviving modes at low energies are the zero-modes. While this is standard for theories with physical degrees of freedom, it is not entirely clear that there are no subtleties for the case at hand, where we are dealing with ``pseudo-forms''---one of which ($B$) has the wrong-sign kinetic term. With that disclaimer, we proceeded and found results that are aligned with expectations. For the circle reduction we arrived at a Maxwell theory that scales like $1/R$. Although in a free theory one can always rescale the fields to change the overall coefficient, this $1/R$ scaling is also consistent with the Legendre transform of the 6D hamiltonian reduced on the circle, if one works with canonically-conjugate pairs. A logically straightforward next step in this direction would be to explicitly extend the analysis to the nonabelian, 3-algebra version of the theory constructed in \cite{Lambert:2019diy}. For the reduction on the Riemann surface, we recovered the expected 4D $\mathcal N=2$ Seiberg--Witten effective action for two sets of abelian gauge fields, subject to a constraint. Perhaps surprisingly, in the special case where the period matrix of the Seiberg--Witten curve $\tau_{\alpha\beta}$ was independent of the Riemann-surface moduli, this constraint related the gauge fields via standard electric-magnetic duality, reminiscent of the work of \cite{Schwarz:1993vs}. Therefore another interesting direction would be to better understand the nature of the constraint and to what extent it encodes information about electric-magnetic duality in general.

Other directions could involve understanding how to couple $H_{(g)}^J$ to a self-dual string, or exploiting the ideas introduced here to write down a four dimensional Maxwell theory that is manifestly invariant under both Lorentz transformations and duality symmetry, along the lines of what happens for the PST formalism; {\it c.f.} \cite{Medina:1997fn}. From a more speculative perspective, it would be very interesting if there existed a nice geometric construction that accommodates ``pseudo-forms'' and explains the properties of $\cal{\tilde{M}}$. This could shed some light on how to couple this theory to gravity.\footnote{Naively, one would only promote the curved metric $g$ to a dynamical field. Making both curved and flat metrics dynamical would result in something akin to a bimetric theory of gravity  \cite{Rosen:1940zza}.}  Moreover, the fact that $H$ and $B$ mix under diffeomorphisms could be due to both originating from the same object in a higher-dimensional theory, after compactification. For example, the idea that the abelian 6D (2,0) theory can be formulated as a 7D Chern--Simons theory has been put forward in \cite{Witten:1996hc} and further utilised in \cite{Belov:2006jd}.

To summarise, the action discussed here is a novel, relatively simple formulation that is consistent with the abelian, low-energy physics of a single M5-brane in M-theory. It has several attractive features:  it is Lorentz and diffeomorphism covariant without introducing a scalar field that ultimately requires some non-vanishing preferred direction---as \emph{e.g.} is the case in the PST formalism. Although we require additional modes with the wrong-sign kinetic terms these can be discarded---effectively set to zero---when one examines the physical degrees of freedom. In addition, it gives a canonical Poisson-bracket structure to the theory on a generic manifold. We hope to continue its investigation in the near future.

\section*{Acknowledgements}

We would like to thank D.~Berman and A.~Sen for useful discussions and comments. E.A. is funded by a Royal Society Research Fellows Enhancement Award RGF/EA/180073. N.L. is funded in part through the STFC grant ST/L000326/1 and would like to thank the Department of Theoretical Physics at CERN for hospitality during the completion of this work. The work of C.P. is supported by a Royal Society University Research Fellowship UF120032 and in part through the STFC grant ST/P000754/1.



\bibliography{Geo_Act}
\bibliographystyle{utphys}

\end{document}